\documentclass[twocolumn]{aastex63}
\pdfoutput=1
\usepackage{graphicx, url}
\usepackage{enumerate, enumitem}
\usepackage{amssymb, amsmath, amsfonts, upgreek}
\usepackage{gensymb}
\usepackage{mathtools}
\usepackage{hyperref}
\usepackage{xspace}
\usepackage{xcolor}
\usepackage{ulem}

\newcommand{\Mjup}{\mbox{$M_{\rm Jup}$}}

\newcommand{\pmoffs}[2]{^{+ #1}_{- #2}}

\newcommand{\hipparcos}{\textsl{Hipparcos}\xspace}
\newcommand{\Gaia}{\textsl{Gaia}\xspace}
\newcommand{\gaia}{\textsl{Gaia}\xspace}

\newcommand{\orbitcodename}{{\tt orvara}\xspace}
\newcommand{\htofcodename}{\texttt{htof}\xspace}

\shorttitle{Misalignment of the HAT-P-11 system}
\shortauthors{An et al.}
\submitjournal{AJ}
\begin{document}

\title{Significant Mutual Inclinations Between the Stellar Spin and the Orbits of Both Planets in the HAT-P-11 System}

\author[0000-0003-0115-547X]{Qier An}
\affiliation{Department of Physics, University of California, Santa Barbara, Santa Barbara, CA 93106, USA}
\affiliation{Department of Physics and Astronomy Johns Hopkins University, Baltimore, MD 21218, USA}

\author[0000-0003-0834-8645]{Tiger Lu}
\affiliation{Department of Astronomy, Yale University, 52 Hillhouse, New Haven, CT 06511, USA}

\author[0000-0003-0168-3010]{G.~Mirek Brandt}
\affiliation{Department of Physics, University of California, Santa Barbara, Santa Barbara, CA 93106, USA}

\author[0000-0003-2630-8073]{Timothy D.~Brandt}
\affiliation{Space Telescope Science Institute, 3700 San Martin Dr., Baltimore, MD 21218, USA}
\affiliation{Department of Physics, University of California, Santa Barbara, Santa Barbara, CA 93106, USA}

\author[0000-0001-8308-0808]{Gongjie Li}
\affiliation{Center for Relativistic Astrophysics, School of Physics, Georgia Institute of Technology, Atlanta GA 30332, USA}

\begin{abstract}
Planet-star obliquity and planet-planet mutual inclination encode a planetary system's dynamical history, but both of their values are hard to measure for misaligned systems with close-in companions. HAT-P-11 is a K4 star with two known planets: a close-in, misaligned super-Neptune with a $\approx$5 day orbit, and an outer super-Jupiter with a $\approx$10 yr orbit. In this work we present a joint orbit fit of the HAT-P-11 system with astrometry and $S$-index corrected radial velocity data. By combining our results with previous constraints on the orientation of the star and the inner planet, we find that all three angular momenta---those of the star, planet b, and planet c---are significantly misaligned.  We confirm the status of planet c as a super-Jupiter, with $2.68\pm0.41\, {M_{\rm Jup}}$ at a semimajor axis of $4.10\pm0.06\,$ au, and planet b's mass of ${M_b\sin{i_b}}=0.074\pm0.004\, {M_{\rm Jup}}$. We present the posterior probability distribution of obliquity between star A and planet c, and mutual inclination between planet b and planet c.

\end{abstract}

\keywords{---}

\section{Introduction}
The main reservoirs of angular momentum in the Solar system---the Sun's spin and the planetary orbits---are all closely aligned \citep{souami2012solar}.  This observational fact has informed theories of the Solar system planets' formation in a stable protoplanetary disk \citep[see a review in][]{solar_disk}.  Exoplanetary systems show a much wider range of planet masses and orbital separations than our own Solar system, from hot Jupiters and mini-Neptunes to packed planetary systems within the orbit of Venus \citep{gillon2017}.  The masses and orbits of these planets may be measured with transits and radial velocities, but stellar obliquities--the angles between the stellar spin and planet orbit--are harder to measure.  We do not know the extent to which a diversity in exoplanetary masses, eccentricities, and periods is matched by a diversity in obliquities.

Tidal forces tend to damp obliquity between planets and their host star, and can efficiently align close-in planets \citep{dawson_tidal}.  Stellar-disk obliquity also tends to damp, which could lead to low obliquities for planets formed in that disk \citep{obliquity_damp, Li16}.  Despite this, moderate degrees of stellar-disk misalignment may be excited, notably in the presence of a binary companion \citep{zanazzi_misalignment}.  A planet that forms in a disk will inherit the angular momentum of its parent disk.  Larger stellar obliquities may be produced either by more violent events like planet-planet scattering \citep{chatterjee2008dynamical,nagasawa2011orbital}, or by ZLK cycles \citep{von_zeipel_1910,lidov_1962,kozai_1962}.  

Projected obliquities have been measured for a modest number of exoplanetary systems, mostly through the Rossiter--McLaughlin effect \citep{rossiter1924detection,mcLaughlin1924some}. As a planet transits different parts of a star spinning edge on, redshifted and blueshifted patches of the stellar surface can be sequentially occulted, leading to slight variations in absorption line profiles.  The Rossiter--McLaughlin effect is only measurable for transiting planets; obliquities of more widely separated planets that are unlikely to transit are much harder to measure. Misalignment between the orbits of different planets (planet-planet mutual inclinations) in an exoplanetary system may sometimes be constrained by transit-timing variations \citep{holman2005, agol2005}, especially when multiple planets transit \citep{obliquity_multiple_planets,2021PSJ.....2....1A}. Multiple transiting planets are likeliest in an aligned system \citep{single_multiple_transit}, and transit-timing variations are most effective at measuring systems with low planet-planet mutual inclinations. The Rossiter--McLaughlin effect was first measured for HAT-P-11 star and the inner planet b by \citet{2010ApJ_Winn_hatp11b_obliquity} and by \citet{2011PASJ_hatp11b_obliquity}. In this work, we use the geometric constraints of planet b's orbit from \citet{2011ApJ_Sanchis-Ojeda_hatp11_star_spin}, in which they combine the previous Rossiter--McLaughlin observations and the spot-crossing anomaly analysis to constrain the inner planet's (planet b) obliquity and the stellar inclination. A similar spot-crossing-anomaly was performed by \citet{2011-hatp11b-transit-Deming}.

Planet-star obliquity and planet-planet mutual inclination encode the dynamical history of a system: they show the final result of dynamical excitation, scattering, and tidal damping.  This full set of obliquity and mutual inclination measurements is very difficult to obtain for an individual system.  Transit-timing variations are most easily measured for closely packed systems in which multiple planets transit.  Multiple transiting planets are likeliest when their orbits are aligned, and dynamically packed systems typically cannot be stable with significant obliquity \citep{albrecht2013low}.  Planet-planet mutual inclinations are very difficult to measure for misaligned systems, while planet-star obliquities may not be measurable for widely separated planets.  We lack a full set of obliquity and mutual inclination constraints for a dynamically hot system. 

The HAT-P-11 system consists of a super-Neptune planet on a $\approx$5 days orbit \citep{Bakos+Torres+Pal+etal_2010} and an outer, super-Jupiter on a $\approx$10 yr orbit \citep{2018AJ_Yee_hatp11_RVS}.  The inner planet (planet b) has a large projected obliquity with the stellar spin \citep{2011ApJ_Sanchis-Ojeda_hatp11_star_spin}.  The outer planet (planet c) has been posited to have a significant inclination relative to the inner planet in order to explain this obliquity \citep{2018AJ_Yee_hatp11_RVS}. \citet{2020MNRAS_hatp11bc_obliquity} derived a bimodal inclination of planet c, favoring a large mutual inclination between the two planets. Those authors did not have conclusive evidence of a large mutual inclination, however, and suggested
that future data from the Gaia satellite \citep{Gaia_General_2016} could offer such evidence.

In this paper we combine radial velocities (RVs), absolute astrometry from Hipparcos and Gaia, and projected obliquity constraints from \cite{2011ApJ_Sanchis-Ojeda_hatp11_star_spin} to constrain the alignment of all three main angular momentum vectors in the HAT-P-11 system.  These include the first measurements of the relative inclinations between the outer planet (planet c) and the star, and significantly improved measurements of the inclination between planet c and planet b. We confirm that there does indeed exist a significant misalignment between the orbits of planet c and b. 

We organize the paper as follows. In Section \ref{sec:data}, we summarize the data collected and adopted for this study. We discuss details of orbit fitting and obliquity/mutual inclination calculation in Section \ref{sec:method}. The orbital solution and obliquity/mutual inclination calculation results are presented in Section \ref{sec:results}. We discuss the results and conclude in Section \ref{sec:discussion}. We investigate the dynamical origin of this misalignment in a follow-up paper \citep[hereafter Paper 2]{lu_dynamics}.

\section{Data}\label{sec:data}
We combine four types of archival data for our analysis of the HAT-P-11 system: RVs and absolute astrometry of the host star, a measurement of the projected obliquity between the host star and the inner planet, planet b, and the edge on inclination of planet b's orbit from its transits.  In this section we summarize each of these sources of data in turn.  Our orbital fit relies on RVs and absolute astrometry, while the projected obliquity/mutual inclination constrains the relative orientation of the three main angular momentum vectors in the system.

We adopt RVs from the High-Resolution \'Echelle Spectrometer \citep[HIRES, ][]{Vogt+Allen+Bigelow+etal_1994} on the Keck telescope; the RVs were tabulated by \cite{2018AJ_Yee_hatp11_RVS}. Many of these RVs were taken as part of the long-running California Planet Search \citep[CPS,][]{Howard+Johnson+Marcy+etal_2010}.  Spectra were taken with the iodine cell in place for wavelength calibration and reduced using the CPS pipeline. \cite{2018AJ_Yee_hatp11_RVS} used RVs taken through 2017 August to discover an outer planet, planet c, and to fit for the orbits of both planet b and planet c. In 2024, the same group extended the RV baseline by more than six years, lending strong support to the existence of planet c and further constraining its orbit \citep{hatp11_RV_2024}. The 2024 data set consists of 261 individual RV measurements from 144 unique nights covering a $\approx$17 yr baseline, from late 2007 to 2024 with a median precision of 1.3\,m\,s$^{-1}$. We use the same subset of RV data as \cite{hatp11_RV_2024} for our orbit fit.

We adopt absolute astrometry of the star HAT-P-11 (=HIP~97657) from the Hipparcos-Gaia Catalog of Accelerations \citep[HGCA,][]{brandt_cross_cal_gaia_2018,2021arXiv_HGCAEDR3}.  The HGCA cross calibrates Hipparcos and Gaia astrometry onto a common reference frame and ensures that the errors are statistically well behaved.  There are three measured proper motions in the HGCA: one from Hipparcos, one from Gaia, and a long-term proper motion from the position shift between Hipparcos and Gaia.  Differences between the three proper motions in the HGCA indicate acceleration in an inertial reference frame.  HAT-P-11 itself has a $\chi^2$ value of 42.5 for a model of constant proper motion, i.e., the long-term proper motion differs from the Gaia proper motion at more than $6\sigma$ significance.  

While HAT-P-11 is accelerating between Hipparcos and Gaia, in both missions it is satisfactorily fit by a sky path assuming constant proper motion.  In the original Hipparcos reduction \citep{HIP_TYCHO_ESA_1997} HAT-P-11 has an F2 value just over 2, indicating a fit $\approx$2$\sigma$ worse than expected with a constant proper motion sky path.  In the Hipparcos rereduction \cite{vanLeeuwen_2007} a constant proper motion sky path fits HAT-P-11's astrometry slightly better than for the average star.  In Gaia EDR3 \citep{Lindegren+Klioner+Hernandez+etal_2020}, the star has a renormalized unit weight error (RUWE) of 0.94, slightly better than a value of 1 that is typical for a single star.\footnote{Gaia EDR3 astrometry is identical to Gaia DR3 astrometry for single-star fits.} Section \ref{eq:gaia_accel} discusses the significance of the fact that HAT-P-11 does not appear in the Gaia DR3 non-single-star catalog \citep{Halbwachs+Pourbaix+Arenou+etal_2023}.

A satisfactory fit to single-star motion in both Hipparcos and Gaia EDR3 indicates an orbital period significantly longer than the $\approx$3 yr baseline of these missions.  Planet c, with its $\approx$10 yr orbit \citep{2018AJ_Yee_hatp11_RVS}, naturally accounts for the discrepancy between the long-term and Gaia proper motions while also explaining the good fits to single star motion within both Hipparcos and Gaia.  In the following section, we fit orbits to the HAT-P-11 system under the hypothesis that planet c is almost entirely responsible for the observed astrometric acceleration.

Our final input measurement for the HAT-P-11 system is the stellar inclination derived by \cite{2011ApJ_Sanchis-Ojeda_hatp11_star_spin}. This measurement is from a combination of radial velocities via the Rossier-McLaughlin effect and modeling of starspots as the planet crosses the active stellar surface.  \cite{2011ApJ_Sanchis-Ojeda_hatp11_star_spin} find a bimodal probability distribution of stellar inclination  of $80^{+4}_{-3}$ degrees (edge on solution) and $168^{+2}_{-5}$ degrees (pole on solution). The measurement of stellar inclination and projected obliquity does not directly constrain our orbital fit, but it does constrain the relative orientation of the three dominant angular momenta in the system: the spin of the star and the orbits of planets b and c.

\subsection{The Magnetic Activity of HAT-P-11}

HAT-P-11 has been reported to have an unusual level of magnetic activity for a planet-hosting K star with a long rotation period \citep{Morris+Hawley+Hebb+etal_2017}.  Those authors reported a value of $\log_{10}R'_{\rm HK}$, roughly the flux in the chromospheric Ca\,{\sc ii} HK emission lines to that in the underlying continuum, of $-4.35$.  \cite{Basilicata+Giacobbe+Bonomo+etal_2024} cited this level of activity, together with a similar period and a similar (but not identical) phase of Ca\,{\sc ii} HK emission line variations and RV variations, to dispute the existence of planet c.  Those authors instead attribute the long-term RV signal to magnetic activity.  Part of the evidence for this comes from a study of stars observed by HARPS \citep{Lovis+Dumusque+Santos+etal_2011} that showed significant long-term RV variability correlated with levels of magnetic activity.  A value of $\log_{10} R'_{\rm HK} = -4.35$ would make HAT-P-11 the most active star in that sample by a significant margin, and would lead to large RV variability.  In this subsection we briefly review the magnetic activity measurement of HAT-P-11 and its status as a magnetically active K dwarf.  

\begin{figure}
    \includegraphics[width=\linewidth]{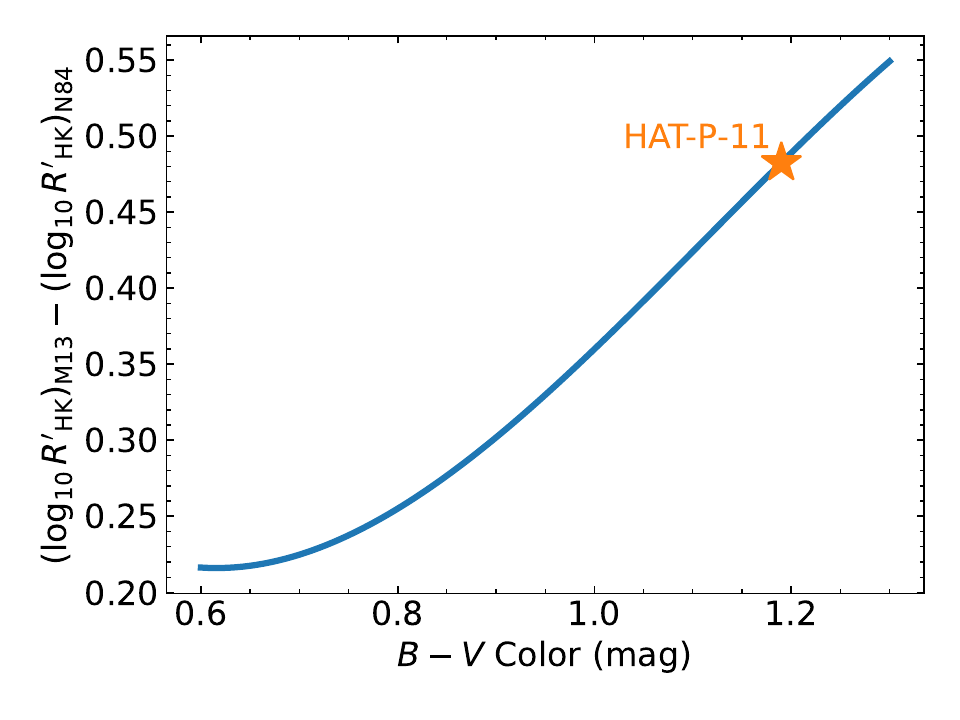}
    \caption{Difference between $\log_{10} R'_{\rm HK}$ computed using the formulas in \cite{Mittag+Schmitt+Schroeder_2013} and \cite{Noyes+Hartmann+Baliunas+etal_1984}, neglecting a correction from photospheric effects (this correction is small for active stars).  The difference between the two $R'_{\rm HK}$ measurements is large and strongly color dependent.  For HAT-P-11, with an $S$-index of 0.58 and a color of $B-V=1.19$ \citep{Morris+Hawley+Hebb+etal_2017}, the \cite{Noyes+Hartmann+Baliunas+etal_1984} and \cite{Mittag+Schmitt+Schroeder_2013} relations give $\log_{10} R'_{\rm HK}$ of $-4.40$ and $-4.82$, respectively, when including the photospheric term. \label{fig:RpHK_compare}}
\end{figure}

Measurements of Ca\,{\sc ii} HK have a long history \citep{Wilson_1963,Wilson_1978}, with the measurements commonly being made using the so-called Mt.~Wilson $S$-index \citep{Vaughan+Preston+Wilson_1978}.  Commonly used conversions to a value known as $R'_{\rm HK}$ were derived and published by \cite{Middelkoop_1982} and \cite{Noyes+Hartmann+Baliunas+etal_1984}.  These converted values of $R'_{\rm HK}$ form the basis for a large body of later work \citep[e.g.][]{Mamajek+Hillenbrand_2008}.  Much more recently, \cite{Mittag+Schmitt+Schroeder_2013} introduced a new conversion between the $S$-index and $R'_{\rm HK}$.  This new conversion differs strongly from the older one published in \cite{Noyes+Hartmann+Baliunas+etal_1984}.  Figure \ref{fig:RpHK_compare} shows the difference between $\log_{10} R'_{\rm HK}$ computed using the two methods, neglecting a correction for photospheric emission (this contribution is small for active stars).  

The literature measurement of $\log_{10} R'_{\rm HK}=-4.35$ from \cite{Morris+Hawley+Hebb+etal_2017} uses the \cite{Mittag+Schmitt+Schroeder_2013} definition of $R'_{\rm HK}$.  Applying Equations (2), (3), (4), (17), (18), and (24) of that paper with $S=0.58$ and $B-V=1.19$ produces $\log_{10} R'_{\rm HK}=-4.40$, but this difference could be due to a slightly different conversion of effective temperature.  The large sample of \cite{Lovis+Dumusque+Santos+etal_2011} instead uses the formulas of \cite{Middelkoop_1982} and \cite{Noyes+Hartmann+Baliunas+etal_1984}; these give a value of $\log_{10} R'_{\rm HK}=-4.82$ for the same $S$-index and $B-V$ color.  Applying the correction in \cite{Lovis+Dumusque+Santos+etal_2011}  for the high metallicity of HAT-P-11 \citep[e.g.][]{Furlan+Ciardi+Cochran+etal_2018} would increase this slightly, to about $\log_{10} R'_{\rm HK}=-4.79$.

A value of $\log_{10} R'_{\rm HK}=-4.79$ would make HAT-P-11's stellar activity typical of mid K stars ($T_{\rm eff} \approx 4700\,{\rm K}$) in the \cite{Lovis+Dumusque+Santos+etal_2011} sample (see their Figure 4).  Figure 6 of \cite{Morris+Hawley+Hebb+etal_2017} confirms that HAT-P-11 has a typical activity level for an old K dwarf.  This modest level of magnetic activity means that the peak-to-valley change in $R'_{\rm HK}$, with $S$ varying from about 0.45 to 0.65 \citep{Morris+Hawley+Hebb+etal_2017}, would be about $0.5 \times 10^{-5}$.  To achieve a peak-to-valley RV change of 60\,m\,s$^{-1}$, this would require a $RV-R'_{\rm HK}$ slope of $\approx$100 in Figure 17 of \cite{Lovis+Dumusque+Santos+etal_2011}, far higher than those authors observed for any other K dwarf (typically $\lesssim 10$).

In summary, HAT-P-11 shows a level of magnetic activity typical of an old K star.  Accounting for the long-term RV variations with magnetic activity would require correlation between RV and Ca\,{\sc ii} S-index with an exceptionally high amplitude \citep{Lovis+Dumusque+Santos+etal_2011}.  With added evidence coming from the $>$$6\sigma$ acceleration in the HGCA, and from extended RV monitoring presented by \cite{hatp11_RV_2024}, we proceed with the hypothesis that the long-term RV variation is due to planet c.


\section{Methodology}\label{sec:method}

\subsection{Orbit Fit}

We fit Keplerian orbital elements to radial velocities and absolute astrometry using the code \textit{orvara} \citet{2021AJ_orvara}. \textit{orvara} uses parallel-tempered Markov chain Monte Carlo (MCMC) with {\tt ptemcee} \citep{Foreman-Mackey+Hogg+Lang+etal_2013,Vousden+Farr+Mandel_2016}. We use a parallel-tempered MCMC with 20 temperatures; for each temperature we use 100 walkers with 400,000 steps per walker, and each walker is thinned by a factor of 400. We use the coldest chain for statistical inference.  Our MCMC chains converged after 100,000 steps; we conservatively discard the first 200,000 as burn in. Convergence is checked individually for each parameter. For a 100 walker 1000 step (thinned) chain, the correlation length is $\leq 5-10$ thinned steps, depending on the parameter.  Thus, our chains have $\gtrsim$10$^4$ quasi-independent points for each parameter. 

We use the standard \textit{orvara} priors for most parameters, as in e.g., \citet{2021AJ_orvara, 2021AJ_Li_RV_planets, 2021AJ_GBrandt2021}.  These standard priors include a geometric prior on inclination ($p(i) = \sin i$), log-uniform prior in semimajor axis, and uniform priors on eccentricity, orbital phase, argument of periastron, and orientation in the plane of the sky.  We adopt an informative prior of $0.81\pm0.03$ for the host star's mass \citep{Dai+Winn+Berta-Thompson+etal_2018} and uniform priors on the masses of the two planets (b \& c): uniform between 0 and 100\,$\Mjup$.  These differ from the log-uniform companion mass priors that \textit{orvara} uses by default. This choice of planet masses priors yields a slightly higher mass compared to the default log-uniform (by much less than one sigma). A summary of priors is indicated in the second column of Table \ref{tab:posterior_params}. 

At each MCMC step, absolute astrometry is processed by \htofcodename \citep{htof_main_paper, htof_zenodo}. Given a set of orbital parameters, \htofcodename models the absolute astrometry that Gaia and Hipparcos would observe.  Using the epoch absolute astrometry itself is not yet possible for Gaia, while these epoch astrometry in Hipparcos are problematic for orbit fitting \citep{Brandt+Michalik+Brandt_2023}.  Instead, \htofcodename uses the epochs and orientations of observations from the \gaia GOST service\footnote{\url{https://gaia.esac.esa.int/gost/}}, and both \hipparcos and \hipparcos 2 (from the 2014 Java Tool IAD but released in 2022) from the European Space Agency archive\footnote{\url{https://www.cosmos.esa.int/web/hipparcos/interactive-data-access}} to compute synthetic epoch astrometry. From these, we obtain synthetic positions and proper motions that can be statistically compared to the HGCA values.  

We use the RV data as updated and extended by \citet{hatp11_RV_2024}, which consist of 261 RV points, to conduct our orbit fit for the HAT-P-11 system. The RV fit includes an analytic marginalization of the unknown RV zero-point, i.e., the RV of the system barycenter.  As described in the following subsection, it also includes an analytic marginalization over a correlation between the measured Ca\,{\sc ii} $S$-index and the measured RV.  

Our {\it orvara} fits to the absolute astrometry and RVs can constrain many, though not all, of the orbital parameters.  The inclination of the inner planet (planet b) is almost entirely unconstrained by {\it orvara}: 
the motion induced by planet b's orbit on the star, the submicroarcsecond semiamplitude signal, is too small to be detectable in either Hipparcos or Gaia. Besides, given planet b's short orbit of $\approx$5 days and \hipparcos and \gaia long baselines of over years, time-averaged astrometric signal are undetectable for planet b. This inclination is instead constrained to be nearly edge on by the light curves of the planetary transits \citep{Bakos+Torres+Pal+etal_2010}. The position angle of planet b's orbit--its orientation in the plane of the sky--is unconstrained by any observations. 
All of the orbital parameters for planet c can be meaningfully constrained in our fit. Table \ref{tab:posterior_params} describes MCMC priors and posterior orbital parameters of planet b and planet c, where the posterior mass distribution for planet b is reported in the form of $\mathrm{M_b}\sin{i_b}$, the lower mass limit, though transit data confirm that $\sin i \approx 1$.

\subsection{S-index Correction}

The Ca\,{\sc ii} HK $S$-index correlates with the strength of magnetic activity, which is in turn correlated with convection and can induce small shifts in measured RVs \citep{Lovis+Dumusque+Santos+etal_2011}.  The original version of {\it orvara} does not include a correlation between stellar activity and RV in the orbit fit. To account for the magnetic activity of HAT-P-11~A, we modified {\it orvara} to marginalize over an $S$-index-RV correlation coefficient.  Th presence of this new parameter is limited to the RV portion of the likelihood (Equation (25) in \cite{2021AJ_orvara}).

The addition of an $S$-index correlation adds one additional free parameter to the likelihood, but it appears linearly in the model of the RVs.  As a result, it may be integrated over in exactly the same way as the RV zero-point in \cite{2021AJ_orvara}.  The marginalization is now an integral over a two-dimensional Gaussian, and involves the computation of the covariance matrix between RV zero-point and $S$-index together with the best-fit values for both.  This calculation is performed, assuming a single RV instrument, using the branch {\tt RVs\_Sindex\_1inst\_analytic}.

We perform orbital fits both with and without fitting for an $S$-index-RV correlation.  The effect on most paramters is small, though the $S$-index does absorb some of the RV variation and produce a slightly lower mass constraint on planet c. The results from both fits are presented in Table \ref{tab:posterior_params}.

\subsection{Obliquity/Mutual Inclination Calculation}

We use the orbital posterior of HAT-P-11~c from our orbit fit to directly calculate the planet's three-dimensional orbital angular momentum vector (note that there turn out to be two possible inclination modes). \citet{2011ApJ_Sanchis-Ojeda_hatp11_star_spin} solved for the posterior of the 3D stellar spin axis, albeit up to one unknown angle, $\lambda$. This is the angle that the spin vector is rotated about the line-of-sight vector to Earth\footnote{$\lambda$ is also commonly referred to as the projected obliquity, when it is in reference not to the line-of-sight vector but the orbital angular momentum vector of a specific planet.}. \citet{2011ApJ_Sanchis-Ojeda_hatp11_star_spin} measured stellar inclination of HAT-P-11, in short, by combining Rossiter--Mclaughlin (RM) measurements of the obliquity between star A and inner planet b (see, \citealp{2011PASJ_hatp11b_obliquity, 2010ApJ_Winn_hatp11b_obliquity}) with Kepler light-curve variability and models of stellar active latitudes and starspot locations. 

To solve for the obliquity between the star A and planet c (hereafter $\psi_{\rm Ac}$), we adopt the marginalized stellar inclination $i_{\rm star}$ posteriors from Table 3 of \citet{2011ApJ_Sanchis-Ojeda_hatp11_star_spin}. The pole on case is $i_{\rm star}=168\degree\,^{+2}_{-5}$ and edge on is $i_{\rm star}=80\degree\,^{+4}_{-3}$. According to \citet{2011ApJ_Sanchis-Ojeda_hatp11_star_spin}, their two predictions of $v\sin{i_{\rm star}} = (2\pi R_*/R_{rot})\sin{i_{\rm star}}$ value equal 0.5 km/s (pole on) and 1.3 km/s (edge on), which each agree with the observed value of $1.5\pm1.5$ km/s from \citet{Bakos+Torres+Pal+etal_2010}. However, modern measurements of $v\sin{i_{\rm star}}$, $2\pm0.5$ km/s from \citet{2018ApJS..237...38B} and $3.2\pm1$ km/s from \citet{2017AJ....154..107P}, slightly favor (by about 2$\sigma$ to 3$\sigma$) the edge on scenario. We include and analyze the edge on and pole on stellar-spin cases separately. 

We have a full 3d stellar spin vector given a stellar inclination $i_{\rm star}$ and a sky-plane angle $\lambda$ which separates the spin and orbital angular momenta vectors. We already have the planet's three-dimensional orbital angular momentum, but $\lambda$ is unknown and can assume any value between 0 and 2$\pi$. Thus to find the obliquity between A and c, we calculate the dot product between the two vectors and marginalize over the uniform distribution of sky-plane angles $\lambda$. The dot product (and thus $\cos{\psi_{\rm Ac}}$) may be calculated with Equation (8) of \citet{2011ApJ_Sanchis-Ojeda_hatp11_star_spin}:
\begin{align*}
    \cos{\psi} = \cos{i_{\rm star}}\cos{i_{\rm orbit}}+\sin{i_{\rm star}}\cos{\lambda}\sin{i_{\rm orbit}}.
\end{align*}
It depends on only the stellar and orbital inclinations ($i_{\rm star}$ and $i_{\rm c}$) and $\lambda$. 

\begin{figure}
    \centering
    \includegraphics[width=0.95\linewidth]{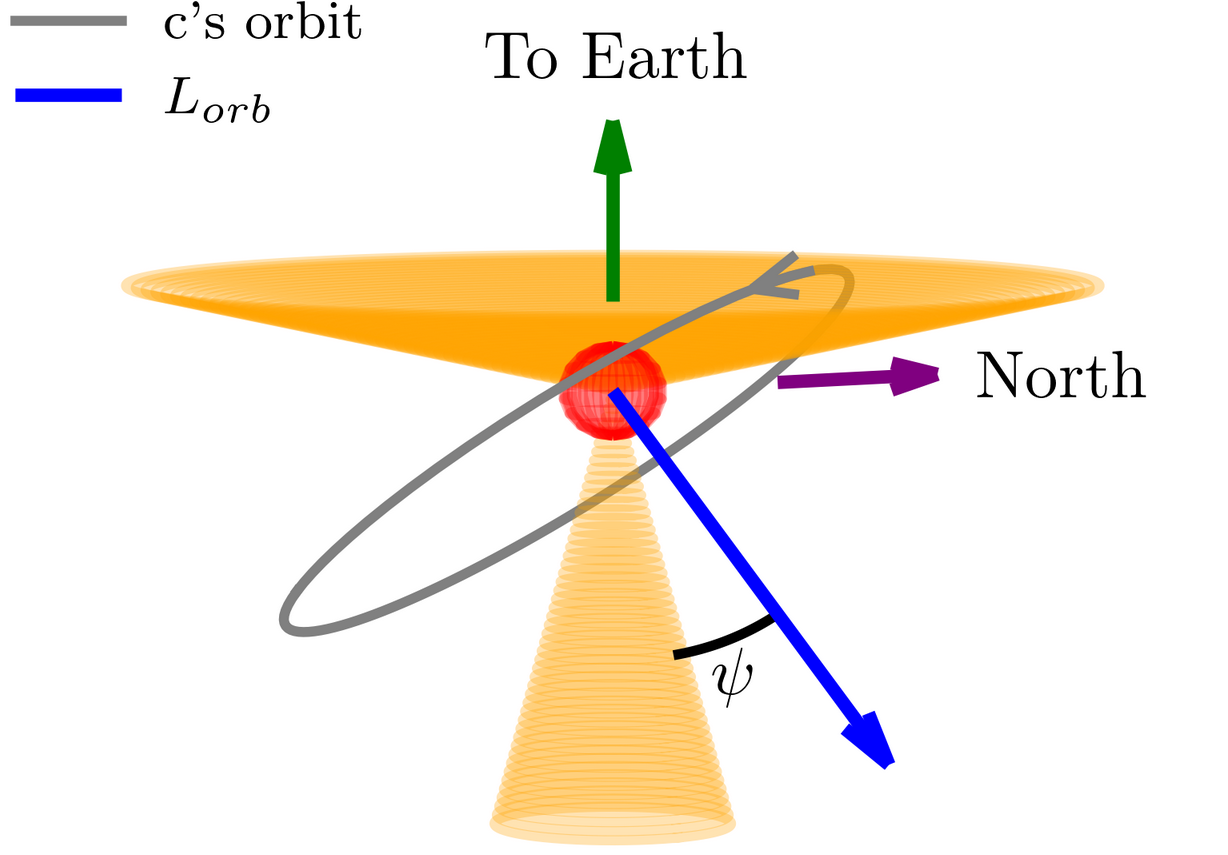}
    \caption{Orbital architecture of the HAT-P-11 system. The star (not to scale) is represented by the red sphere. North and line-of-sight (to Earth) are given. The orbit of c is described in grey, while the inner planet b is not shown. The two orange cones represent the possible stellar spins that \citet{2011ApJ_Sanchis-Ojeda_hatp11_star_spin} derived from obliquity measurements of b. The upper cone is the set of edge on stellar-spin configurations and the bottom cone give the pole on solutions. The blue vector is the orbital angular momentum of the outer planet c.}
    \label{fig:3d}
\end{figure}

The orbital architecture is plotted in Figure \ref{fig:3d}, where the orange cones are sets of possible stellar spins (upper cone for edge on and bottom cone for pole on), and the blue vector shows planet c's orbital angular momentum. The cosine of the obliquity between planet c and the star A is the dot product of $L_{\rm orb}$ with all vectors living on the two spin cones. One possible obliquity is indicated by $\psi$, which is roughly 20\degree.

We solve for the mutual inclination between planet b and c ($i_{\rm mut, bc}$) in an identical fashion, only with $i_{\rm b}$ taking the place of $i_{\rm star}$, and $i_{\rm b}$ being constrained by the planet's transit light curves. Since we do not have a full 3d normal vector of planet b's orbit, we have no constrain on the sky-plane angle $\lambda$. Therefore, in $i_{\rm mut, bc}$ calculation, we assume $\lambda$ can be any value between 0 and $2\pi$. Including \hipparcos-\gaia accelerations in the three-body fit breaks the $m\sin{i}$ degeneracy of the long-period outer planet, but the accelerations do not constrain the inclination $i_{\rm b}$ of the inner planet due to that planet's tiny mass and short period, and its resulting sub-$\mu$as astrometric orbital amplitude. We use $i_{\rm b} = 88.\!\!\degree5 \pm 0.\!\!\degree6$ from \citet{2017AJ_hatp11_b_inclination}, who measured it from the transiting planet's impact parameter, and draw values of $\lambda$ from a uniform distribution from 0 to $2\pi$.

\section{Results}\label{sec:results}

\begin{deluxetable*}{lrrrrr}
\tablecaption{MCMC prior and posterior orbital parameters for HAT-P-11~b and HAT-P-11~c, with and without S-index correction. \label{tab:posterior_params} }
\tablehead{
\colhead{Parameter} & \colhead{Prior} & \colhead{Posterior~b$\pm 1\sigma$\tablenotemark{$*$}} & \colhead{Posterior~b$\pm 1\sigma$} & \colhead{Posterior~c$\pm 1\sigma$\tablenotemark{$*$}} &
\colhead{Posterior~c$\pm 1\sigma$}
}
\startdata
Jit (m/s)    &  $1/\sigma_{\rm Jit}$ &  
${5.43}_{-0.34}^{+0.37}$ & ${4.07}_{-0.20}^{+0.21}$ & 
${5.43}_{-0.34}^{+0.37}$ & ${4.07}_{-0.20}^{+0.21}$   \\
$\mathrm{M_{pri}}$ ($\mathrm{M_\odot}$)  & N(0.81, 0.03) &     ${0.811\pm0.03}$ &${0.811\pm0.03}$ &     ${0.811\pm0.03}$ &${0.811\pm0.03}$ \\
$\mathrm{M_{sec}}$($\mathrm{M_{jup}}$) & U(0, 100)  &  $0.073\pm0.004$\tablenotemark{$a$} & $0.074\pm0.004$\tablenotemark{$a$} &  $3.06\pm0.042$ & ${2.68\pm0.41}$ \\
a (AU)    &   $1/a$ & ${0.05258\pm0.0006}$ & ${0.05258\pm0.0006}$ &  
${4.19\pm0.07}$ & 
${4.10\pm0.06}$  \\
$\sqrt{\varepsilon}\sin{\omega}$   &  U($-$1, 1) & ${0.16}_{-0.16}^{+0.13}$& ${0.23}_{-0.10}^{+0.09}$  & ${0.492\pm0.044}$ & ${0.495\pm0.029}$  \\
$\sqrt{\varepsilon}\cos{\omega}$   & U($-$1, 1) &  ${0.36}_{-0.15}^{+0.10}$ &  ${0.44}_{-0.06}^{+0.05}$  & ${-0.563\pm0.039}$ & ${-0.637\pm0.023}$   \\
Mean longitude $\lambda_{ref}$ ($\degree$)\tablenotemark{$b$}    & U($-$180, 180) & ${79.4\pm4.4}$ & ${86.4\pm2.8}$& ${327.2\pm3.5}$ &  ${323.4\pm1.0}$  \\
Period P(days)  &  \nodata     & ${4.8877\pm0.0002}$ &  ${4.8880\pm0.0001}$ & ${3474}_{-62}^{+69}$ & ${3361} \pm 31$ \\
argument of periastron $\omega$ ($\degree$) & \nodata   &    ${34}_{-21}^{+111}$ & ${28}_{-11}^{+11}$ &
${138.9\pm4.1}$ & ${142.1\pm2.5}$ \\
eccentricity $e$  &   \nodata & ${0.171}_{-0.075}^{+0.068}$& ${0.251}_{-0.047}^{+0.045}$&   
${0.560\pm0.036}$ & ${0.652\pm0.017}$  \\
\hline
inclination $i$ (\degree) (mode 1) & $\sin i$   &  \nodata &  \nodata &  ${33.5}_{-4.4}^{+6.1}$ & ${33.1}_{-4.9}^{+7.3}$  \\
Ascending node $\Omega$ ($\degree$) (mode 1) & U($-$180, 180) & \nodata  &  \nodata & ${117.1.1\pm7.9}$ &  ${109.1\pm8.9}$   \\
inclination i (\degree) (mode 2)  & $\sin i$  &   \nodata &  \nodata & ${143.6}_{-6.2}^{+4.7}$ & ${145.8}_{-7.3}^{+5.1}$   \\
Ascending node $\Omega$ ($\degree$) (mode 2) & U($-$180, 180) & \nodata  &   \nodata & ${40\pm10}$ & $46\pm10$   \\
\hline
$\orbitcodename$ Reference Epoch ${(t_{\rm ref})}$ & 2455197.50 BJD & \nodata & \nodata & \nodata & \nodata 
\enddata
\tablenotetext{*}{Results without S-index correction}
\tablenotetext{a}{$\mathrm{M_{b}}$$\mathrm{sin(i)}$ constraint}
\tablenotemark{b}
{reference epoch at 2010.0} 
    \tablecomments{Orbital elements all refer to orbit of the companion about the barycenter. The orbital parameters for the primary about each companion are identical except $\omega_{A} = \omega + \pi$. We use $\pm \sigma$ to denote the $1\sigma$ Gaussian error about the median when the posteriors are approximately symmetric. Otherwise, we denote the value by median$\pmoffs{u}{l}$ where $u$ and $l$ denote the  68.3\% confidence interval about the median. The reference epoch $t_{\rm ref}$ is not a fitted parameter and has no significance within the fit itself, it is the epoch at which the Mean Longitude $(\lambda_{\rm ref})$ is evaluated. Mode 1 refers to the orbital parameters derived if one selects the inclination mode with $i<90$, and mode 2 the converse. The two $i$/$\Omega$ modes yield orbits with otherwise identical parameters (masses, semi-major axis, etc..); visually the orbits are mirror images of each other in 3D space.}
\end{deluxetable*}

\begin{figure}
    \centering    \includegraphics[width=0.48\textwidth]{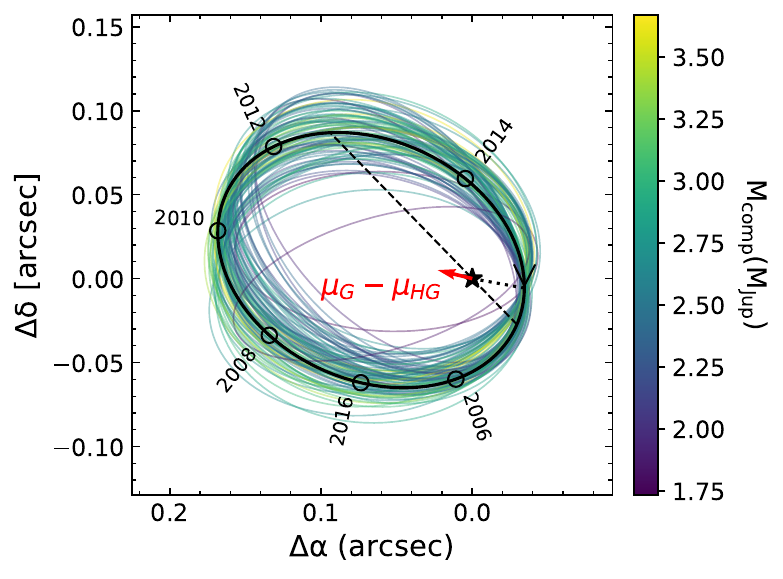} \hspace{3pt} \\
    \hspace{7pt} \includegraphics[width=0.45\textwidth]{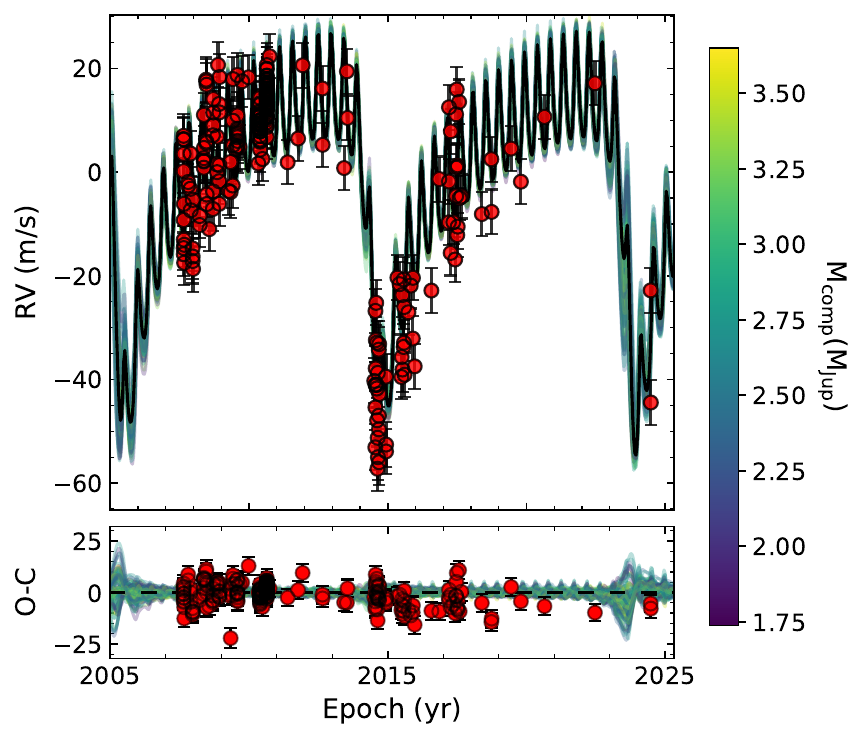}
    \caption{\textbf{Top:} relative orbits of planet c (in arc seconds) from fitting RVs and absolute astrometry from the v.EDR3 HGCA. 100 random orbital draws are shown and color coded by companion mass. Positions at a few selected epochs are displayed for the maximum-likelihood orbital solution (black). The host star is marked by the star symbol at the origin. The change of proper motion from Hipparcos-Gaia to Gaia is indicated as a red vector. \textbf{Bottom:} The observed RVs of HAT-P-11 overplotted with the best-fit orbit (in black) and a random sampling of other orbits from the MCMC chain. The RV signal includes both the high-frequency oscillations from b and the long-term high-eccentricity undulation of c. The RV residuals with respect to the best-fit orbit are in the attached bottom subpanel.}
    \label{fig:astrometric_orbits_c}
\end{figure}

\begin{figure*}
    \centering
    \includegraphics[width=0.9\linewidth]{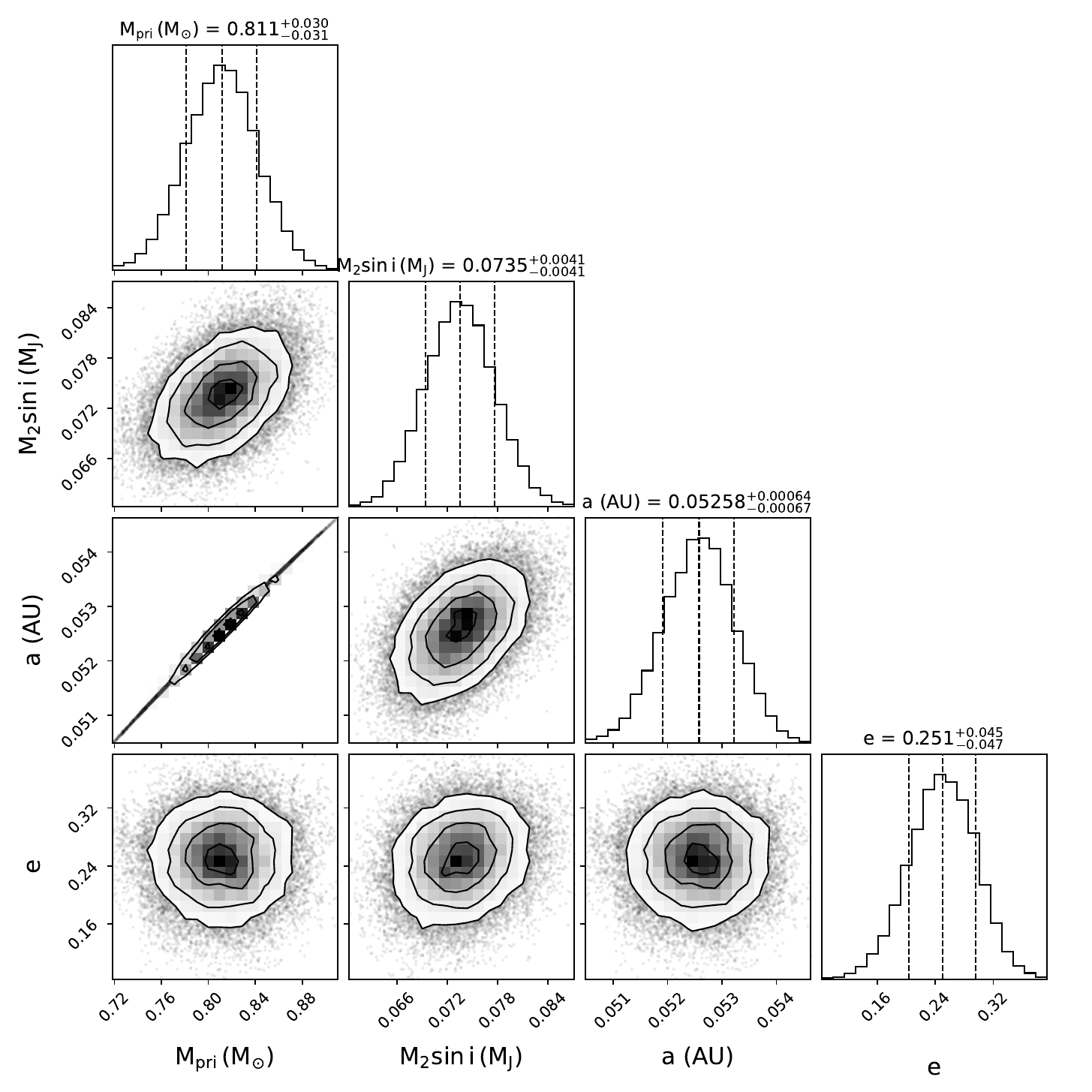}
    \caption{Corner plot of the orbital elements, with respect to the star, for HAT-P-11~b, which demonstrating a constraint on $\mathrm{M_b}\sin{i}$. In the 1D histograms, the vertical-dashed lines about the center dashed lines give the $16\%$ and $84\%$ quantiles around the median. In the 2d histograms, the contours give the 1$\sigma$, 2$\sigma$, and 3$\sigma$ levels. The summary statistics of HAT-P-11 system orbit fit are given separately for both inclination modes in Table \ref{tab:posterior_params}.}
    \label{fig:corner-b}
\end{figure*}

Here we present the results of a three-body orbit fit of the HAT-P-11 system, as well as the obliquity measurements between the primary star HAT-P-11 A and its outer planet c ($\psi_{\rm Ac}$), and the mutual inclination between inner planet b and outer planet c ($i_{\rm mut, bc}$). A summary of the prior and best-fit posterior orbital parameters of the HAT-P-11 system is reported in Table \ref{tab:posterior_params}, where we report the orbit solution with and without the S-index-RV correction of {\it orvara}. The top panel of Figure \ref{fig:astrometric_orbits_c} presents relative orbits of the outer planet c drawn from the MCMC posterior samples, together with the best-fit orbit in black.  The bottom panel of Figure \ref{fig:astrometric_orbits_c} shows the observed RVs of the primary star together with random orbits drawn from the MCMC posterior and the best-fit orbit in black. 
We present the corner plot of posterior orbital elements in Figure \ref{fig:corner-b} for planet b and Figure \ref{fig:corner-c} for planet c. Figure \ref{fig:inc_modes} presents the astrometric orbit of HAT-P-11 due to planet c. Figures \ref{fig:obliquities}-\ref{fig:obliquities_b_c}
report the probability distribution of obliquity between star A and planet c, and mutual inclination between planet b and planet c.

\subsection{Orbital Elements of the Planets}

RVs constrain the orbital period, eccentricity, and $M \sin i$ of the inner planet, planet b.  Our mass, eccentricity, and period agree with those originally determined from transits \citep{Bakos+Torres+Pal+etal_2010}.  The inner planet is responsible for the high-frequency RV signal visible in Figure \ref{fig:astrometric_orbits_c}.  The longer-term variations are due to planet c, whose orbital results we now briefly discuss.

The parameters of planet c are constrained jointly by absolute astrometry and RV data. The presence of an outer planet c is inferred from astrometric acceleration of HAT-P-11 A and the long-term undulation of the RV curve presented in the lower panel of Figure \ref{fig:astrometric_orbits_c}; the planet was first identified and fit by \cite{2018AJ_Yee_hatp11_RVS} using only RVs. 
We measure a mass of $2.68\pm0.41~\mathrm{M_{jup}}$ for planet c, confirming its status as a super-Jupiter. Our result of $\mathrm{M_c}$ is $\approx$1$\sigma$ heavier than the previous value of $2.3_{-0.5}^{+0.7} ~ \mathrm{M_{jup}}$ from \citet{2020MNRAS_hatp11bc_obliquity}. That analysis used a previous, less-precise version of Hipparcos-Gaia astrometry based on Gaia DR2 \citep{brandt_cross_cal_gaia_2018}. 
Our semi-major axis of $4.10\pm0.06$ au agrees with \citet{2018AJ_Yee_hatp11_RVS}'s finding of $4.13^{+0.29}_{-0.16}$ au. Our result also characterizes the orbit with argument of periastron ($\omega = 142.\!\!\degree1\pm2.\!\!\degree5$) and eccentricity ($e = 0.652\pm0.017$). Our inclination posterior is bimodal, where the peaks of the distribution are close to $30\degree$ and $150\degree$.

\begin{figure*}
    \centering
    \includegraphics[width=\linewidth]{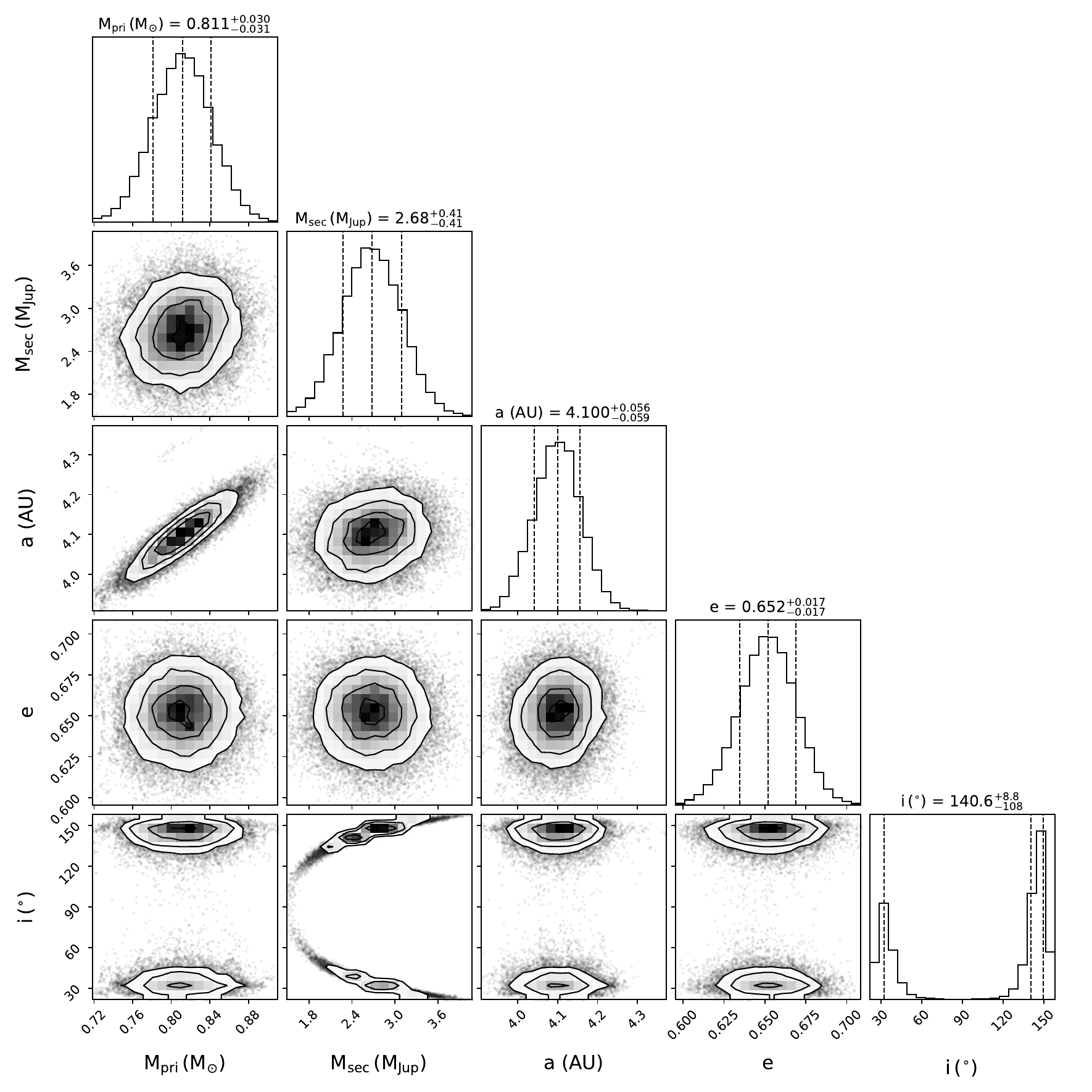}
    \caption{Corner plot of the orbital elements, with respect to the star, for HAT-P-11~c. In the 1D histograms, the vertical-dashed lines about the center dashed lines give the 16\% and 84\% quantiles around the median. In the 2d histograms, the contours give the 1$\sigma$, 2$\sigma$, and 3$\sigma$ levels. The result of inclination is bimodal (shown in lower-right 1D histogram), where the peaks of distribution are close to $30\degree$ and $150\degree$ (two modes are reflected with respect to $90\degree$). As Figure \ref{fig:inc_modes} shows, the retrograde mode ($i > 90$) is slightly favored by the Hipparcos proper motion. The summary statistics of HAT-P-11 system orbit fit are given separately for both inclination modes in Table \ref{tab:posterior_params}.}
    \label{fig:corner-c}
\end{figure*}

\begin{figure}
    \centering
    \includegraphics[width=0.48\textwidth]{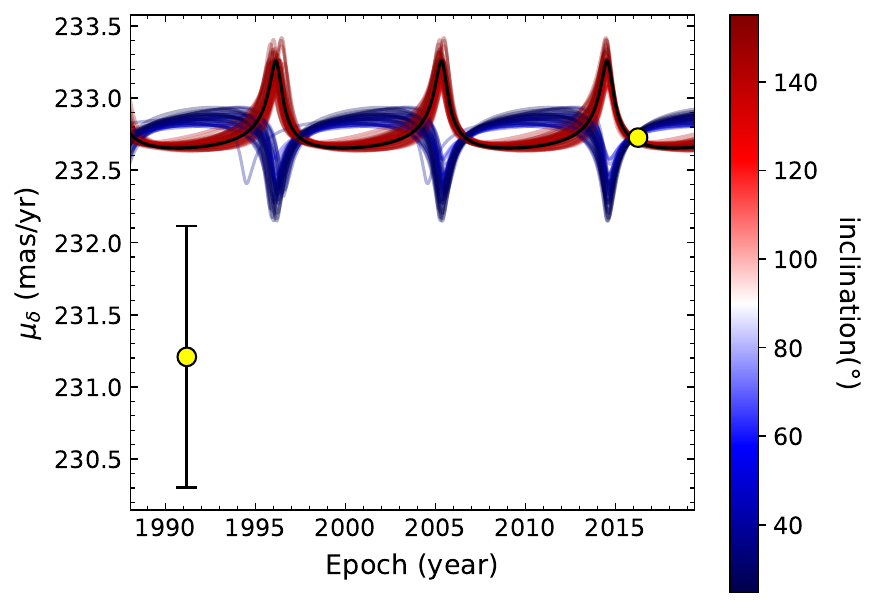}
    \caption{Astrometric orbit of HAT-P-11 due to planet c. One hundred random orbital draws are shown and color coded by inclination; the maximum-likelihood orbit is plotted in black. Retrograde orbits (red) are slightly favored by the Hipparcos proper motion (yellow point with large error bar on left), and both modes fits well with Gaia proper motion (yellow point on the right, with an error bar smaller than the point size.)}
    \label{fig:inc_modes}
\end{figure}

The upper panel of Figure \ref{fig:astrometric_orbits_c} shows the orbit of planet c in the plane of the sky.  The maximum-likelihood orbit is colored black, while the colored lines are randomly chosen from the orbit fitting posteriors. The change of proper motion $\mu_{\rm G}-\mu_{\rm HG}$ is indicated as a red vector, which is scaled to be visible in this plot.

We present the corner plot of planet c's posterior orbital elements in Figure \ref{fig:corner-c}. The smooth, nearly Gaussian contours indicate a well-converged posterior. Most plotted parameters are well constrained except for inclination, in which two modes are reflected with respect to $90\degree$. The degeneracy of inclination can be visualized in the bottom row of plots and in the lower-right 1D histogram. Figure \ref{fig:inc_modes} presents the astrometric orbit of HAT-P-11 due to planet c. The orbits are plotted with respect to time (year) and proper motion in declination $\mu_\delta$ (mas/yr). The fitted orbits are drawn from our result chain and color coded by inclination. Prograde orbits are colored blue and retrograde orbits are colored red; the maximum-likelihood orbit is plotted in black. The retrograde mode ($i > 90$) is slightly favored by the Hipparcos proper motion (yellow point with large error bar on left), and both modes fits well with Gaia proper motion (yellow point on the right, with error bar smaller than point size.) The degeneracy of inclination is common when Hipparcos and Gaia provide only two effective proper motion measurements (with the Hipparcos proper motion being too imprecise to matter).  Two proper motions plus an RV curve are insufficient to distinguish prograde from retrograde orbits \citep{Kervella+Arenou+Schneider_2020}.  

The dynamical mass, semi-major axis, and eccentricity of planet c are not significantly affected by the inclination degeneracy.
The obliquity between the star A and planet c ($\psi_{\rm Ac}$), however, is affected.  We therefore treat the two modes separately in the following obliquity calculation. We present the posterior probability distribution of obliquity between the star A and planet c as well as the mutual inclination between planet b and planet c in the following sections. 

\subsection{Obliquity Between Star A and Planet c}

\begin{figure}
    \centering    \includegraphics[width=0.99\linewidth]{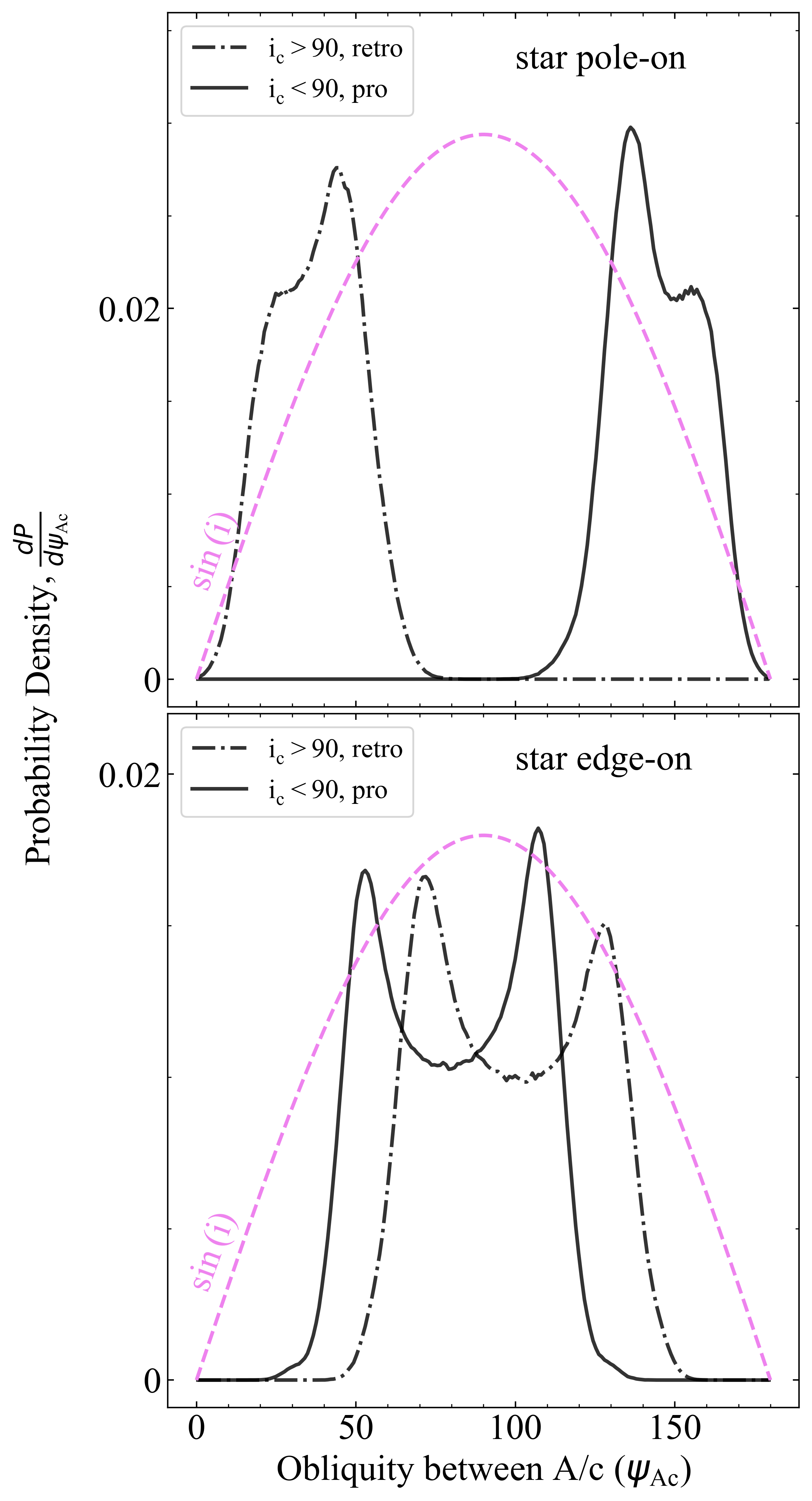}
    \caption{The obliquity posterior between the outer planet HAT-P-11~c and the star A, for the four possible orbital/stellar spin configurations. The top panel shows the two obliquity posteriors (one for each of the possible inclination modes of the planet) assuming the pole on stellar inclination posterior from \citet{2011ApJ_Sanchis-Ojeda_hatp11_star_spin}. The bottom panel uses the the edge on stellar inclination posterior. In dashed violet is the geometric $\sin{i}$ posterior, which would be expected if we had no constraint on the orbit of HAT-P-11~c.}
    \label{fig:obliquities}
\end{figure}

We separately calculate the obliquity between star A and planet c ($\psi_{\rm Ac}$) for two scenarios according to possible stellar spin from \citet{2011ApJ_Sanchis-Ojeda_hatp11_star_spin}: star pole on ($i_{\rm star}=168\degree\,^{+2}_{-5}$) and star edge on ($i_{\rm star}=80\degree\,^{+4}_{-3}$). In each scenario, $\psi_{\rm{Ac}}$ is then calculated separately for two orbital modes: retrograde mode ($i_{\rm c}>90\degree$) and prograde mode($i_{\rm c}<90\degree$). The obliquity posterior between star A and the outer planet c is plotted in Figure \ref{fig:obliquities} for the four possible orbital/stellar spin configurations. The upper and bottom panels show the probability density of $\psi_{\rm Ac}$ for star pole on and edge on scenarios. These probability densities would not be flat in the absence of constraining data, as two randomly oriented vectors will tend to be closer to perpendicular than to being parallel.  This imposes a geometric $\sin\psi$ prior, which we plot as a dashed-violet line for comparison. Deviations from this $\sin \psi$ prior reflect the constraining power of our measurement.  

\begin{figure}
    \centering  \includegraphics[width=0.9\linewidth]{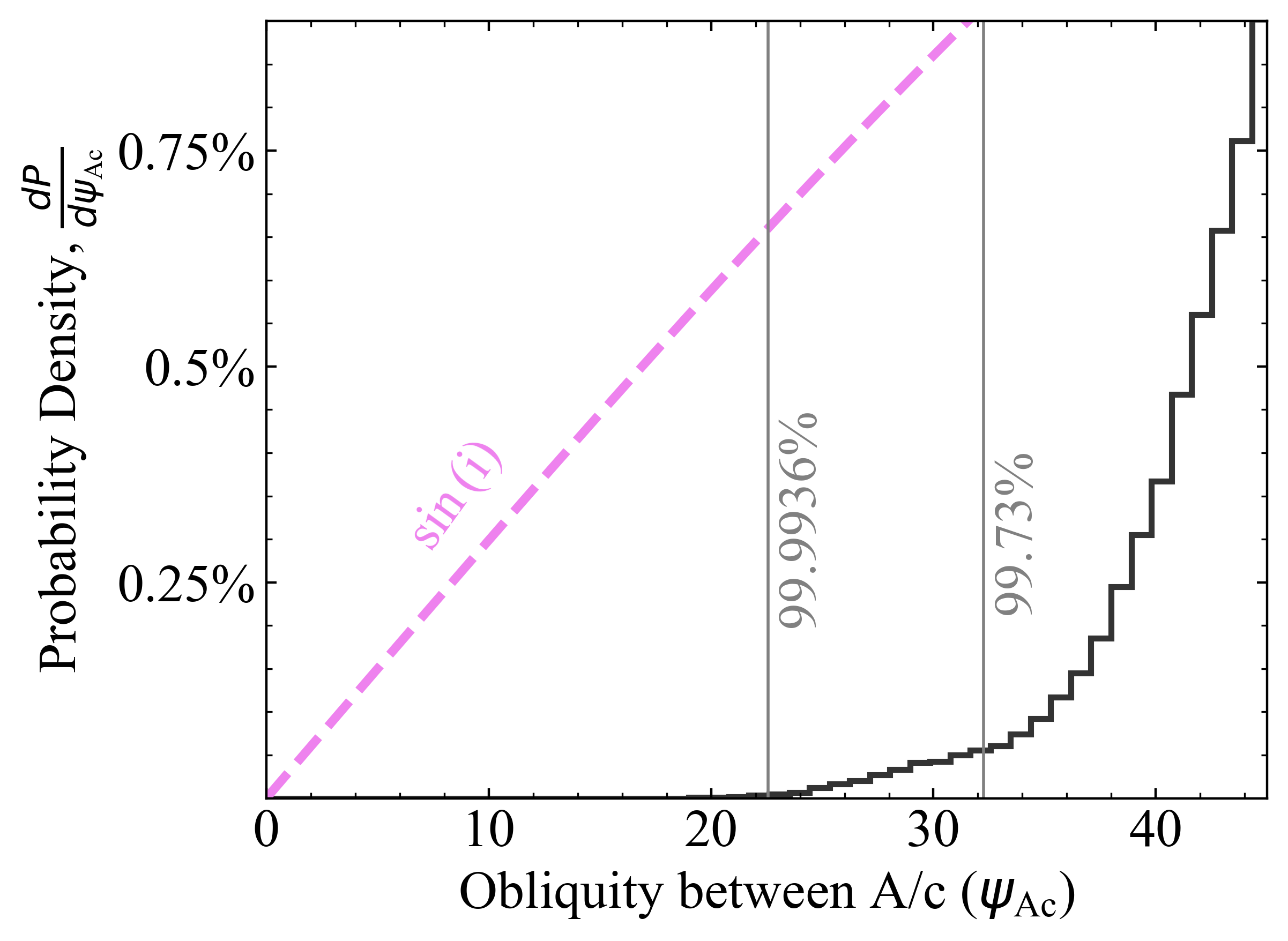}
    \caption{Zoom in of Figure \ref{fig:obliquities} between 0\degree and 40\degree obliquities. We show the most likely case (edge on) and the $i_{\rm c}<90$\degree orbital solution. Two lower limits are presented as  gray vertical lines: the $3\sigma$ lower-limit on the obliquity is 32\degree, and the $4\sigma$ (99.9936\%) lower-limit is 22\degree.}
    \label{fig:obliquities_zoom}
\end{figure}

Our posterior on the obliquity between the star and planet c strongly disfavors alignment.  Figure \ref{fig:obliquities_zoom} shows a zoom-in of the $\psi_{\rm Ac}$ posterior in Figure \ref{fig:obliquities} for the case where the star appears edge on. Obliquities less than $32\degree$ can be ruled out with 99.73\% confidence, and obliquities less than 22\degree are disfavored by odds of least 15,000-to-1 (i.e., 4$\sigma$ confidence).

In all the cases, but especially if HAT-P-11 A is seen edge on, $\psi_{\rm Ac}$ near 0$\degree$ are disfavored significantly more than from the $\sin{i}$ prior alone. In the star pole on scenario, $\psi_{\rm{Ac}}>90\degree$ and and $\psi_{\rm{Ac}}<90\degree$ is strongly ruled out for $i_{\rm c}>90\degree$ and $i_{\rm c}<90\degree$, respectively. The peak value of $\psi_{\rm{Ac}}$ is around $50\degree$ for retrograde mode and around $150\degree$ for prograde mode. In the star edge on scenario, $\psi_{\rm{Ac}}<30\degree$ and $\psi_{\rm{Ac}}>150\degree$ are strongly ruled out for both inclination modes. The maximum value of $\psi_{\rm{Ac}}$ is around $80\degree$ and the second maximum value is around $135\degree$ for retrograde mode. The maximum is around $110\degree$ and the second maximum is around $55\degree$ for prograde mode.  These local maxima extend less than a factor of 2 above the continuum of the probability distribution.  While we can determine that planet c's orbit is misaligned with respect to the stellar spin we obtain only a broad constraint on the angle of the misalignment.

\subsection{Mutual Inclination Between Planets b and c}

\begin{figure}
    \centering    \includegraphics[width=0.99\linewidth]{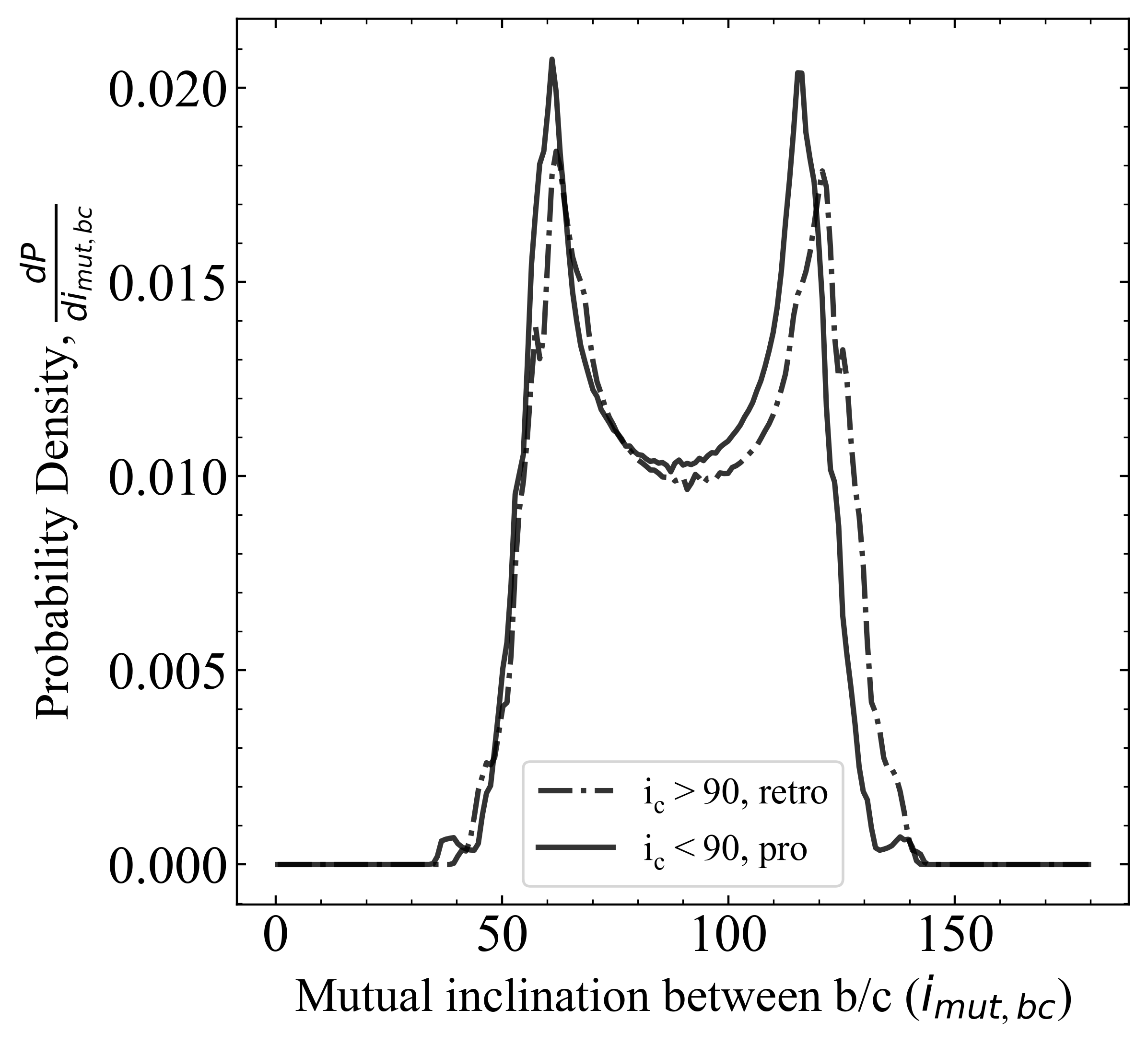}
    \caption{The mutual inclination posterior between the outer c and the inner planet b, for the two possible $i_c$ configurations, assuming $i_b = 88.5\degree\pm0.6$ from \citep{2017AJ_hatp11_b_inclination}.}
    \label{fig:obliquities_b_c}
\end{figure}

The mutual inclination between the inner planet b and the outer planet c ($i_{\rm mut, bc}$) is also calculated separately for the two $i_{\rm c}$ modes. Due to its close-in orbit and extremely short period, we are not able to constrain planet b's orbital inclination with a combination of absolute astrometry and RV data. Instead, we use the well-determined inclination of $i_{\rm b} = 88.\!\!\degree5\pm0.\!\!\degree6$ for inner planet b derived from the transit light curve and reported by \citet{2017AJ_hatp11_b_inclination}. 

Figure \ref{fig:obliquities_b_c} presents the mutual inclination posterior between the outer planet c and the inner planet b. The peaks are at 59$.\!\!\degree$7 and 116$.\!\!\degree$7 for planet c prograde mode, and at 61$.\!\!\degree$5 and 120$.\!\!\degree$3 for planet c retrograde mode.
The distribution of $i_{\rm mut, bc}$ is slightly broader for the retrograde mode than for the prograde mode. Comparing to the previous study from \citet{2020MNRAS_hatp11bc_obliquity}, this work treats the two inclination modes of planet c individually, and obtains an improved posterior distribution of $i_{\rm mut, bc}$ (see the right panel of their Figure 2). The maximum-likelihood peaks from this work, with combined inclination modes of planet c, are at 58$.\!\!\degree$8 and 117$.\!\!\degree$6, which are larger than both maximum-likelihood values from \citet{2020MNRAS_hatp11bc_obliquity}. The posterior distribution from this work strongly rules out $i_{\rm mut, bc}<30\degree$ and $i_{\rm mut, bc}>150\degree$, providing stronger evidence for misalignment.

\subsection{Gaia Acceleration \label{eq:gaia_accel}}

The bimodal inclination distribution suggests that, based on the existing data, it is impossible to distinguish whether planet c's orbit is prograde or retrograde. Future astrometric data from \Gaia would help break this degeneracy.

To measure the impact of Gaia on our knowledge of planet c's orbit, we predict what Gaia DR4 could report for the acceleration terms on RA ($\alpha$) and Dec ($\delta$) using \textit{htof} \citep{htof} and the \Gaia GOST scanning predictions. We treat the two inclination modes of planet c individually for acceleration calculations. These calculations simulate absolute astrometry based on an orbital configuration from our joint posteriors, and then use a seven-parameter fit (linear motion plus constant acceleration) to obtain a Gaia prediction.  

\begin{figure}
    \centering    \includegraphics[width=0.95\linewidth]{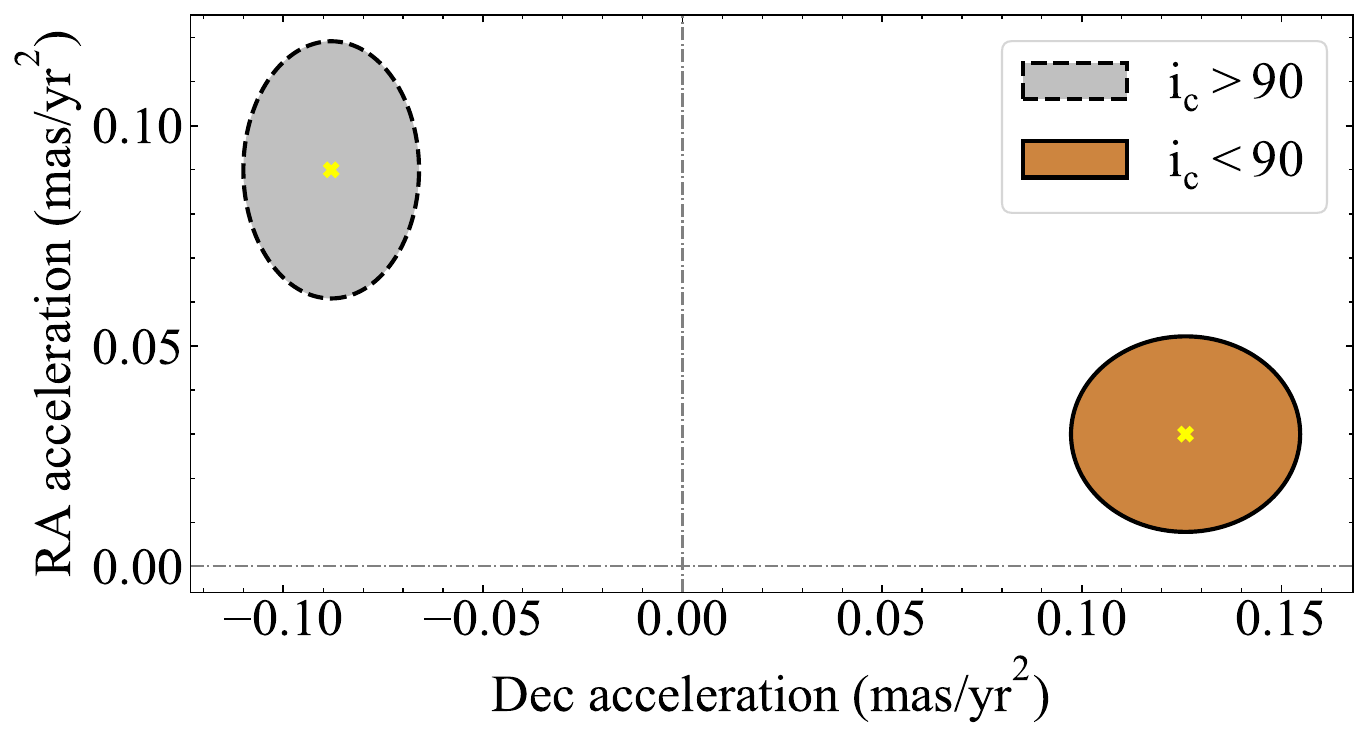}
    \caption{The predicted accelerations for two inclination modes of planet c, in unit of milli-arcsecond per year squared. The $1\sigma$ contour of the retrograde mode is plotted with a dashed line, and the corresponding contour of prograde mode is shown with a solid line. The center yellow crosses indicate the mean values. Zero accelerations in both RA and Dec are marked for reference.}
    \label{fig:accel}
\end{figure}

Figure \ref{fig:accel} plots the acceleration posteriors for both inclination modes of planet c's orbit. A yellow cross is plotted at each ellipse indicates the mean value of acceleration, where the area of each ellipse include the $1\sigma$ uncertainty in the prediction. These modes will be readily distinguishable by Gaia assuming an acceleration precision better than $\approx$0.1\,mas\,yr$^{-2}$.  Most acceleration measurements in Gaia DR3 already exceed this precision \citep{Halbwachs+Pourbaix+Arenou+etal_2023}, while sensitivity to acceleration is a steep function of mission duration thanks to the quadratic dependence of the accumulated signal on time.  It appears likely that the next Gaia data release will settle the 3D orientation of planet c's orbit.

Figure \ref{fig:accel} raises the question of why Gaia DR3 has not already measured the astrometric acceleration of HAT-P-11.  While astrometric accelerations are commonly measured to a precision better than 0.1\,mas\,yr$^{-2}$ in DR3, the smallest reported accelerations in 7-parameter acceleration fits are $\approx$0.7\,mas\,yr$^{-2}$, a factor of $\approx$5 smaller than our prediction for HAT-P-11.  Stars with astrometric accelerations of this magnitude (e.g.~Gaia DR3 2085682921206479488) have renormalized unit weight errors $\approx$1.5, just above the conservative threshold of 1.4 to accept a non-single-star solution \citep{Halbwachs+Pourbaix+Arenou+etal_2023}, and have parallax precisions comparable to that of HAT-P-11.  With its much lower predicted acceleration, it is unsurprising that HAT-P-11 does not appear in the Gaia DR3 non-single star catalog. 

\section{Discussion and Conclusions}\label{sec:discussion}

Through a combination of absolute astrometry, radial velocity, and transit-derived measurements, we confirm that the three main angular momentum vectors in the HAT-P-11 system are all significantly misaligned.  Our constraints on the degree of the misalignment depend on the stellar inclination, which \cite{2011ApJ_Sanchis-Ojeda_hatp11_star_spin} determined to be consistent with either a nearly pole on or a nearly edge on orientation for the star. 

The case where the star is viewed edge on produces a greater misalignment (a minimum misalignment of $15\degree$ versus a minimum of $\approx5\degree$ at $4\sigma$ significance), however edge on is also a more likely scenario for the star. It is geometrically favored, as visualized in Figure \ref{fig:3d}, by comparing the solid-angle captured by the upper, edge on cone to the much smaller solid-angle subtended by the lower pole on cone. Secondly, the edge on solution produces a rotational velocity more favored by modern measurements. \citet{2011ApJ_Sanchis-Ojeda_hatp11_star_spin} stated that the pole on solution predicted a $v\sin{i}$ for the star of $0.5$\,km/s, while edge on was consistent with $1.3$\,km/s. At the time, both were consistent with the $v\sin{i}$ measurements available, but as of this publication that is no longer the case. \citet{2018ApJS..237...38B} and \citet{2017AJ....154..107P} measured $v\sin{i} =$ $2\pm 0.5$\,km/s and $3.2\pm 1$\,km/s, respectively, favoring the edge on scenario. The edge on spin case is thus favored by about 2$\sigma$ to 3$\sigma$ considering the relative disagreement with $v\sin{i}$ measurements; and favored by a factor of about 30 geometrically. \citet{2017-Morris-hatp11-starspots} rejected the pole on solution, because the $\approx3\%$ rotational variability would not be caused by the high-latitude spots viewed from a star pole on orientation, which agrees with our conclusion here.

\begin{deluxetable}{l|l|l}
\tablecaption{Confidence interval of derived obliquity between A/c and mutual inclination between b/c with combined $i_c$ modes, assuming the star to rotate edge on as seen from Earth.}
\label{Tab:confidence_interval}
\tablewidth{0pt}
\tablehead{Parameter & 99\% confidence & 95\% confidence}
\startdata
$\psi_{\rm Ac}$ & 33$.\!\!\degree$3 to 147$.\!\!\degree$0  & 44$.\!\!\degree$7 to 138$.\!\!\degree$4\\ 
\hline
${i_{\rm mut, bc}}$ & 38$.\!\!\degree$5 to 140$.\!\!\degree$0 & 49$.\!\!\degree$0 to 131$.\!\!\degree$3
\enddata
\end{deluxetable}

Table \ref{Tab:confidence_interval} presents the 99\% and 95\% confidence intervals of obliquity and mutual inclination posteriors from this work, combining the two inclination modes for planet c ($i_{\rm c}$), assuming the star to rotate edge on as seen from Earth. At $2\sigma$ significance, the three angular momentum vectors are all misaligned to at least $\approx$40$\degree$, while at $3\sigma$ significance, they are all misaligned to at least $\approx$30$\degree$. A companion paper, Paper 2, explores the dynamical origin and history of the HAT-P-11 system.

This work, by using the improvements of Gaia EDR3 \citep{2020GaiaEDR3_catalog_summary,Lindegren+Klioner+Hernandez+etal_2020}, has significantly improved our knowledge of the three-dimensional structure of the HAT-P-11 system. We present the first constraints on the obliquity between the star and the outer planet, planet c $\psi_{\rm Ac}$, and present an improved $i_{\rm mut, bc}$ posterior, indicating a high mutual inclinations within HAT-P-11 system and suggesting a complex dynamical history. Based on the results of this study, Paper 2 explores the origin and evolution of the HAT-P-11 system through a combination of scattering and subsequent tides and migration.


\acknowledgments
We thank the anonymous referee for a very helpful and constructive report. TDB and QA gratefully acknowledge funding from NASA XRP through grant \#80NSSC21K0574. GL is grateful for the support by NASA 80NSSC20K0641 and 80NSSC20K0522. This work makes use of data from Hipparcos-Gaia Catalog of Accelerations \citep[HGCA,][]{brandt_cross_cal_gaia_2018}, the High-Resolution \'Echelle Spectrometer \citep[HIRES, ][]{Vogt+Allen+Bigelow+etal_1994} on the Keck telescope, the long-running California Planet Search \citep[CPS,][]{Howard+Johnson+Marcy+etal_2010}, and the RVs tabulated by \cite{2018AJ_Yee_hatp11_RVS} and \cite{hatp11_RV_2024}. 

This work uses astropy \citep{astropy:2013, astropy:2018}, scipy \citep{2020SciPy-NMeth}, numpy \citep{numpy1, numpy2}, \htofcodename \citep{htof_zenodo, htof_main_paper}, orvara \citep{2021AJ_orvara}, and Jupyter (\url{https://jupyter.org/}).

\bibliographystyle{aasjournal}
\bibliography{refs.bib}

\begin{thebibliography}{}
\expandafter\ifx\csname natexlab\endcsname\relax\def\natexlab#1{#1}\fi
\providecommand{\url}[1]{\href{#1}{#1}}
\providecommand{\dodoi}[1]{doi:~\href{http://doi.org/#1}{\nolinkurl{#1}}}
\providecommand{\doeprint}[1]{\href{http://ascl.net/#1}{\nolinkurl{http://ascl.net/#1}}}
\providecommand{\doarXiv}[1]{\href{https://arxiv.org/abs/#1}{\nolinkurl{https://arxiv.org/abs/#1}}}

\bibitem[{{Agol} {et~al.}(2005){Agol}, {Steffen}, {Sari}, \& {Clarkson}}]{agol2005}
{Agol}, E., {Steffen}, J., {Sari}, R., \& {Clarkson}, W. 2005, \mnras, 359, 567, \dodoi{10.1111/j.1365-2966.2005.08922.x}

\bibitem[{{Agol} {et~al.}(2021){Agol}, {Dorn}, {Grimm}, {Turbet}, {Ducrot}, {Delrez}, {Gillon}, {Demory}, {Burdanov}, {Barkaoui}, {Benkhaldoun}, {Bolmont}, {Burgasser}, {Carey}, {de Wit}, {Fabrycky}, {Foreman-Mackey}, {Haldemann}, {Hernandez}, {Ingalls}, {Jehin}, {Langford}, {Leconte}, {Lederer}, {Luger}, {Malhotra}, {Meadows}, {Morris}, {Pozuelos}, {Queloz}, {Raymond}, {Selsis}, {Sestovic}, {Triaud}, \& {Van Grootel}}]{2021PSJ.....2....1A}
{Agol}, E., {Dorn}, C., {Grimm}, S.~L., {et~al.} 2021, PSJ, 2, 1, \dodoi{10.3847/PSJ/abd022}

\bibitem[{Albrecht {et~al.}(2013)Albrecht, Winn, Marcy, Howard, Isaacson, \& Johnson}]{albrecht2013low}
Albrecht, S., Winn, J.~N., Marcy, G.~W., {et~al.} 2013, The Astrophysical Journal, 771, 11

\bibitem[{{Astropy Collaboration} {et~al.}(2013){Astropy Collaboration}, {Robitaille}, {Tollerud}, {Greenfield}, {Droettboom}, {Bray}, {Aldcroft}, {Davis}, {Ginsburg}, {Price-Whelan}, {Kerzendorf}, {Conley}, {Crighton}, {Barbary}, {Muna}, {Ferguson}, {Grollier}, {Parikh}, {Nair}, {Unther}, {Deil}, {Woillez}, {Conseil}, {Kramer}, {Turner}, {Singer}, {Fox}, {Weaver}, {Zabalza}, {Edwards}, {Azalee Bostroem}, {Burke}, {Casey}, {Crawford}, {Dencheva}, {Ely}, {Jenness}, {Labrie}, {Lim}, {Pierfederici}, {Pontzen}, {Ptak}, {Refsdal}, {Servillat}, \& {Streicher}}]{astropy:2013}
{Astropy Collaboration}, {Robitaille}, T.~P., {Tollerud}, E.~J., {et~al.} 2013, \aap, 558, A33, \dodoi{10.1051/0004-6361/201322068}

\bibitem[{{Bakos} {et~al.}(2010){Bakos}, {Torres}, {P{\'a}l}, {Hartman}, {Kov{\'a}cs}, {Noyes}, {Latham}, {Sasselov}, {Sip{\H{o}}cz}, {Esquerdo}, {Fischer}, {Johnson}, {Marcy}, {Butler}, {Isaacson}, {Howard}, {Vogt}, {Kov{\'a}cs}, {Fernandez}, {Mo{\'o}r}, {Stefanik}, {L{\'a}z{\'a}r}, {Papp}, \& {S{\'a}ri}}]{Bakos+Torres+Pal+etal_2010}
{Bakos}, G.~{\'A}., {Torres}, G., {P{\'a}l}, A., {et~al.} 2010, \apj, 710, 1724, \dodoi{10.1088/0004-637X/710/2/1724}

\bibitem[{{Basilicata} {et~al.}(2024){Basilicata}, {Giacobbe}, {Bonomo}, {Scandariato}, {Brogi}, {Singh}, {Di Paola}, {Mancini}, {Sozzetti}, {Lanza}, {Cubillos}, {Damasso}, {Desidera}, {Biazzo}, {Bignamini}, {Borsa}, {Cabona}, {Carleo}, {Ghedina}, {Guilluy}, {Maggio}, {Mainella}, {Micela}, {Molinari}, {Molinaro}, {Nardiello}, {Pedani}, {Pino}, {Poretti}, {Southworth}, {Stangret}, \& {Turrini}}]{Basilicata+Giacobbe+Bonomo+etal_2024}
{Basilicata}, M., {Giacobbe}, P., {Bonomo}, A.~S., {et~al.} 2024, \aap, 686, A127, \dodoi{10.1051/0004-6361/202347659}

\bibitem[{{Brandt} {et~al.}(2021{\natexlab{a}}){Brandt}, {Brandt}, {Dupuy}, {Li}, \& {Michalik}}]{2021AJ_GBrandt2021}
{Brandt}, G.~M., {Brandt}, T.~D., {Dupuy}, T.~J., {Li}, Y., \& {Michalik}, D. 2021{\natexlab{a}}, \aj, 161, 179, \dodoi{10.3847/1538-3881/abdc2e}

\bibitem[{Brandt \& Michalik(2020)}]{htof_zenodo}
Brandt, G.~M., \& Michalik, D. 2020, Zenodo, \dodoi{10.5281/zenodo.4118572}

\bibitem[{{Brandt} {et~al.}(2023){Brandt}, {Michalik}, \& {Brandt}}]{Brandt+Michalik+Brandt_2023}
{Brandt}, G.~M., {Michalik}, D., \& {Brandt}, T.~D. 2023, RAS Techniques and Instruments, 2, 218, \dodoi{10.1093/rasti/rzad011}

\bibitem[{{Brandt} {et~al.}(2021{\natexlab{b}}){Brandt}, {Michalik}, {Brandt}, {Li}, {Dupuy}, \& {Zeng}}]{htof_main_paper}
{Brandt}, G.~M., {Michalik}, D., {Brandt}, T.~D., {et~al.} 2021{\natexlab{b}}, \aj, 162, 230, \dodoi{10.3847/1538-3881/ac12d0}

\bibitem[{{Brandt} {et~al.}(2021{\natexlab{c}}){Brandt}, {Michalik}, {Brandt}, {Li}, {Dupuy}, \& {Zeng}}]{htof}
---. 2021{\natexlab{c}}, \aj, 162, 230, \dodoi{10.3847/1538-3881/ac12d0}

\bibitem[{{Brandt}(2018)}]{brandt_cross_cal_gaia_2018}
{Brandt}, T.~D. 2018, The Astrophysical Journal Supplement Series, 239, 31, \dodoi{10.3847/1538-4365/aaec06}

\bibitem[{{Brandt}(2021)}]{2021arXiv_HGCAEDR3}
---. 2021, \apjs, 254, 42, \dodoi{10.3847/1538-4365/abf93c}

\bibitem[{{Brandt} {et~al.}(2021{\natexlab{d}}){Brandt}, {Dupuy}, {Li}, {Brandt}, {Zeng}, {Michalik}, {Bardalez Gagliuffi}, \& {Raposo-Pulido}}]{2021AJ_orvara}
{Brandt}, T.~D., {Dupuy}, T.~J., {Li}, Y., {et~al.} 2021{\natexlab{d}}, \aj, 162, 186, \dodoi{10.3847/1538-3881/ac042e}

\bibitem[{{Brewer} \& {Fischer}(2018)}]{2018ApJS..237...38B}
{Brewer}, J.~M., \& {Fischer}, D.~A. 2018, \apjs, 237, 38, \dodoi{10.3847/1538-4365/aad501}

\bibitem[{Chatterjee {et~al.}(2008)Chatterjee, Ford, Matsumura, \& Rasio}]{chatterjee2008dynamical}
Chatterjee, S., Ford, E.~B., Matsumura, S., \& Rasio, F.~A. 2008, The Astrophysical Journal, 686, 580

\bibitem[{{Dai} {et~al.}(2018){Dai}, {Winn}, {Berta-Thompson}, {Sanchis-Ojeda}, \& {Albrecht}}]{Dai+Winn+Berta-Thompson+etal_2018}
{Dai}, F., {Winn}, J.~N., {Berta-Thompson}, Z., {Sanchis-Ojeda}, R., \& {Albrecht}, S. 2018, \aj, 155, 177, \dodoi{10.3847/1538-3881/aab618}

\bibitem[{{Dawson}(2014)}]{dawson_tidal}
{Dawson}, R.~I. 2014, \apjl, 790, L31, \dodoi{10.1088/2041-8205/790/2/L31}

\bibitem[{{Deming} {et~al.}(2011){Deming}, {Sada}, {Jackson}, {Peterson}, {Agol}, {Knutson}, {Jennings}, {Haase}, \& {Bays}}]{2011-hatp11b-transit-Deming}
{Deming}, D., {Sada}, P.~V., {Jackson}, B., {et~al.} 2011, \apj, 740, 33, \dodoi{10.1088/0004-637X/740/1/33}

\bibitem[{{Foreman-Mackey} {et~al.}(2013){Foreman-Mackey}, {Hogg}, {Lang}, \& {Goodman}}]{Foreman-Mackey+Hogg+Lang+etal_2013}
{Foreman-Mackey}, D., {Hogg}, D.~W., {Lang}, D., \& {Goodman}, J. 2013, \pasp, 125, 306, \dodoi{10.1086/670067}

\bibitem[{{Furlan} {et~al.}(2018){Furlan}, {Ciardi}, {Cochran}, {Everett}, {Latham}, {Marcy}, {Buchhave}, {Endl}, {Isaacson}, {Petigura}, {Gautier}, {Huber}, {Bieryla}, {Borucki}, {Brugamyer}, {Caldwell}, {Cochran}, {Howard}, {Howell}, {Johnson}, {MacQueen}, {Quinn}, {Robertson}, {Mathur}, \& {Batalha}}]{Furlan+Ciardi+Cochran+etal_2018}
{Furlan}, E., {Ciardi}, D.~R., {Cochran}, W.~D., {et~al.} 2018, \apj, 861, 149, \dodoi{10.3847/1538-4357/aaca34}

\bibitem[{{Gaia Collaboration} {et~al.}(2016){Gaia Collaboration}, {Prusti}, {de Bruijne}, {Brown}, {Vallenari}, {Babusiaux}, {Bailer-Jones}, {Bastian}, {Biermann}, {Evans}, {Eyer}, {Jansen}, {Jordi}, {Klioner}, {Lammers}, {Lindegren}, {Luri}, {Mignard}, {Milligan}, {Panem}, {Poinsignon}, {Pourbaix}, {Randich}, {Sarri}, {Sartoretti}, {Siddiqui}, {Soubiran}, {Valette}, {van Leeuwen}, {Walton}, {Aerts}, {Arenou}, {Cropper}, {Drimmel}, {H{\o}g}, {Katz}, {Lattanzi}, {O'Mullane}, {Grebel}, {Holland}, {Huc}, {Passot}, {Bramante}, {Cacciari}, {Casta{\~n}eda}, {Chaoul}, {Cheek}, {De Angeli}, {Fabricius}, {Guerra}, {Hern{\'a}ndez}, {Jean-Antoine-Piccolo}, {Masana}, {Messineo}, {Mowlavi}, {Nienartowicz}, {Ord{\'o}{\~n}ez- Blanco}, {Panuzzo}, {Portell}, {Richards}, {Riello}, {Seabroke}, {Tanga}, {Th{\'e}venin}, {Torra}, {Els}, {Gracia- Abril}, {Comoretto}, {Garcia-Reinaldos}, {Lock}, {Mercier}, {Altmann}, {Andrae}, {Astraatmadja}, {Bellas-Velidis}, {Benson}, {Berthier}, {Blomme}, {Busso}, {Carry}, {Cellino},
  {Clementini}, {Cowell}, {Creevey}, {Cuypers}, {Davidson}, {De Ridder}, {de Torres}, {Delchambre}, {Dell'Oro}, {Ducourant}, {Fr{\'e}mat}, {Garc{\'\i}a-Torres}, {Gosset}, {Halbwachs}, {Hambly}, {Harrison}, {Hauser}, {Hestroffer}, {Hodgkin}, {Huckle}, {Hutton}, {Jasniewicz}, {Jordan}, {Kontizas}, {Korn}, {Lanzafame}, {Manteiga}, {Moitinho}, {Muinonen}, {Osinde}, {Pancino}, {Pauwels}, {Petit}, {Recio-Blanco}, {Robin}, {Sarro}, {Siopis}, {Smith}, {Smith}, {Sozzetti}, {Thuillot}, {van Reeven}, {Viala}, {Abbas}, {Abreu Aramburu}, {Accart}, {Aguado}, {Allan}, {Allasia}, {Altavilla}, {{\'A}lvarez}, {Alves}, {Anderson}, {Andrei}, {Anglada Varela}, {Antiche}, {Antoja}, {Ant{\'o}n}, {Arcay}, {Atzei}, {Ayache}, {Bach}, {Baker}, {Balaguer-N{\'u}{\~n}ez}, {Barache}, {Barata}, {Barbier}, {Barblan}, {Baroni}, {Barrado y Navascu{\'e}s}, {Barros}, {Barstow}, {Becciani}, {Bellazzini}, {Bellei}, {Bello Garc{\'\i}a}, {Belokurov}, {Bendjoya}, {Berihuete}, {Bianchi}, {Bienaym{\'e}}, {Billebaud}, {Blagorodnova}, {Blanco-Cuaresma},
  {Boch}, {Bombrun}, {Borrachero}, {Bouquillon}, {Bourda}, {Bouy}, {Bragaglia}, {Breddels}, {Brouillet}, {Br{\"u}semeister}, {Bucciarelli}, {Budnik}, {Burgess}, {Burgon}, {Burlacu}, {Busonero}, {Buzzi}, {Caffau}, {Cambras}, {Campbell}, {Cancelliere}, {Cantat-Gaudin}, {Carlucci}, {Carrasco}, {Castellani}, {Charlot}, {Charnas}, {Charvet}, {Chassat}, {Chiavassa}, {Clotet}, {Cocozza}, {Collins}, {Collins}, {Costigan}, {Crifo}, {Cross}, {Crosta}, {Crowley}, {Dafonte}, {Damerdji}, {Dapergolas}, {David}, {David}, {De Cat}, {de Felice}, {de Laverny}, {De Luise}, {De March}, {de Martino}, {de Souza}, {Debosscher}, {del Pozo}, {Delbo}, {Delgado}, {Delgado}, {di Marco}, {Di Matteo}, {Diakite}, {Distefano}, {Dolding}, {Dos Anjos}, {Drazinos}, {Dur{\'a}n}, {Dzigan}, {Ecale}, {Edvardsson}, {Enke}, {Erdmann}, {Escolar}, {Espina}, {Evans}, {Eynard Bontemps}, {Fabre}, {Fabrizio}, {Faigler}, {Falc{\~a}o}, {Farr{\`a}s Casas}, {Faye}, {Federici}, {Fedorets}, {Fern{\'a}ndez-Hern{\'a}ndez}, {Fernique}, {Fienga}, {Figueras},
  {Filippi}, {Findeisen}, {Fonti}, {Fouesneau}, {Fraile}, {Fraser}, {Fuchs}, {Furnell}, {Gai}, {Galleti}, {Galluccio}, {Garabato}, {Garc{\'\i}a-Sedano}, {Gar{\'e}}, {Garofalo}, {Garralda}, {Gavras}, {Gerssen}, {Geyer}, {Gilmore}, {Girona}, {Giuffrida}, {Gomes}, {Gonz{\'a}lez-Marcos}, {Gonz{\'a}lez-N{\'u}{\~n}ez}, {Gonz{\'a}lez-Vidal}, {Granvik}, {Guerrier}, {Guillout}, {Guiraud}, {G{\'u}rpide}, {Guti{\'e}rrez-S{\'a}nchez}, {Guy}, {Haigron}, {Hatzidimitriou}, {Haywood}, {Heiter}, {Helmi}, {Hobbs}, {Hofmann}, {Holl}, {Holland}, {Hunt}, {Hypki}, {Icardi}, {Irwin}, {Jevardat de Fombelle}, {Jofr{\'e}}, {Jonker}, {Jorissen}, {Julbe}, {Karampelas}, {Kochoska}, {Kohley}, {Kolenberg}, {Kontizas}, {Koposov}, {Kordopatis}, {Koubsky}, {Kowalczyk}, {Krone-Martins}, {Kudryashova}, {Kull}, {Bachchan}, {Lacoste-Seris}, {Lanza}, {Lavigne}, {Le Poncin-Lafitte}, {Lebreton}, {Lebzelter}, {Leccia}, {Leclerc}, {Lecoeur-Taibi}, {Lemaitre}, {Lenhardt}, {Leroux}, {Liao}, {Licata}, {Lindstr{\o}m}, {Lister}, {Livanou}, {Lobel},
  {L{\"o}ffler}, {L{\'o}pez}, {Lopez-Lozano}, {Lorenz}, {Loureiro}, {MacDonald}, {Magalh{\~a}es Fernandes}, {Managau}, {Mann}, {Mantelet}, {Marchal}, {Marchant}, {Marconi}, {Marie}, {Marinoni}, {Marrese}, {Marschalk{\'o}}, {Marshall}, {Mart{\'\i}n-Fleitas}, {Martino}, {Mary}, {Matijevi{\v{c}}}, {Mazeh}, {McMillan}, {Messina}, {Mestre}, {Michalik}, {Millar}, {Miranda}, {Molina}, {Molinaro}, {Molinaro}, {Moln{\'a}r}, {Moniez}, {Montegriffo}, {Monteiro}, {Mor}, {Mora}, {Morbidelli}, {Morel}, {Morgenthaler}, {Morley}, {Morris}, {Mulone}, {Muraveva}, {Musella}, {Narbonne}, {Nelemans}, {Nicastro}, {Noval}, {Ord{\'e}novic}, {Ordieres-Mer{\'e}}, {Osborne}, {Pagani}, {Pagano}, {Pailler}, {Palacin}, {Palaversa}, {Parsons}, {Paulsen}, {Pecoraro}, {Pedrosa}, {Pentik{\"a}inen}, {Pereira}, {Pichon}, {Piersimoni}, {Pineau}, {Plachy}, {Plum}, {Poujoulet}, {Pr{\v{s}}a}, {Pulone}, {Ragaini}, {Rago}, {Rambaux}, {Ramos-Lerate}, {Ranalli}, {Rauw}, {Read}, {Regibo}, {Renk}, {Reyl{\'e}}, {Ribeiro}, {Rimoldini}, {Ripepi}, {Riva},
  {Rixon}, {Roelens}, {Romero-G{\'o}mez}, {Rowell}, {Royer}, {Rudolph}, {Ruiz-Dern}, {Sadowski}, {Sagrist{\`a} Sell{\'e}s}, {Sahlmann}, {Salgado}, {Salguero}, {Sarasso}, {Savietto}, {Schnorhk}, {Schultheis}, {Sciacca}, {Segol}, {Segovia}, {Segransan}, {Serpell}, {Shih}, {Smareglia}, {Smart}, {Smith}, {Solano}, {Solitro}, {Sordo}, {Soria Nieto}, {Souchay}, {Spagna}, {Spoto}, {Stampa}, {Steele}, {Steidelm{\"u}ller}, {Stephenson}, {Stoev}, {Suess}, {S{\"u}veges}, {Surdej}, {Szabados}, {Szegedi-Elek}, {Tapiador}, {Taris}, {Tauran}, {Taylor}, {Teixeira}, {Terrett}, {Tingley}, {Trager}, {Turon}, {Ulla}, {Utrilla}, {Valentini}, {van Elteren}, {Van Hemelryck}, {van Leeuwen}, {Varadi}, {Vecchiato}, {Veljanoski}, {Via}, {Vicente}, {Vogt}, {Voss}, {Votruba}, {Voutsinas}, {Walmsley}, {Weiler}, {Weingrill}, {Werner}, {Wevers}, {Whitehead}, {Wyrzykowski}, {Yoldas}, {{\v{Z}}erjal}, {Zucker}, {Zurbach}, {Zwitter}, {Alecu}, {Allen}, {Allende Prieto}, {Amorim}, {Anglada-Escud{\'e}}, {Arsenijevic}, {Azaz}, {Balm}, {Beck},
  {Bernstein}, {Bigot}, {Bijaoui}, {Blasco}, {Bonfigli}, {Bono}, {Boudreault}, {Bressan}, {Brown}, {Brunet}, {Bunclark}, {Buonanno}, {Butkevich}, {Carret}, {Carrion}, {Chemin}, {Ch{\'e}reau}, {Corcione}, {Darmigny}, {de Boer}, {de Teodoro}, {de Zeeuw}, {Delle Luche}, {Domingues}, {Dubath}, {Fodor}, {Fr{\'e}zouls}, {Fries}, {Fustes}, {Fyfe}, {Gallardo}, {Gallegos}, {Gardiol}, {Gebran}, {Gomboc}, {G{\'o}mez}, {Grux}, {Gueguen}, {Heyrovsky}, {Hoar}, {Iannicola}, {Isasi Parache}, {Janotto}, {Joliet}, {Jonckheere}, {Keil}, {Kim}, {Klagyivik}, {Klar}, {Knude}, {Kochukhov}, {Kolka}, {Kos}, {Kutka}, {Lainey}, {LeBouquin}, {Liu}, {Loreggia}, {Makarov}, {Marseille}, {Martayan}, {Martinez-Rubi}, {Massart}, {Meynadier}, {Mignot}, {Munari}, {Nguyen}, {Nordlander}, {Ocvirk}, {O'Flaherty}, {Olias Sanz}, {Ortiz}, {Osorio}, {Oszkiewicz}, {Ouzounis}, {Palmer}, {Park}, {Pasquato}, {Peltzer}, {Peralta}, {P{\'e}turaud}, {Pieniluoma}, {Pigozzi}, {Poels}, {Prat}, {Prod'homme}, {Raison}, {Rebordao}, {Risquez}, {Rocca-Volmerange},
  {Rosen}, {Ruiz-Fuertes}, {Russo}, {Sembay}, {Serraller Vizcaino}, {Short}, {Siebert}, {Silva}, {Sinachopoulos}, {Slezak}, {Soffel}, {Sosnowska}, {Strai{\v{z}}ys}, {ter Linden}, {Terrell}, {Theil}, {Tiede}, {Troisi}, {Tsalmantza}, {Tur}, {Vaccari}, {Vachier}, {Valles}, {Van Hamme}, {Veltz}, {Virtanen}, {Wallut}, {Wichmann}, {Wilkinson}, {Ziaeepour}, \& {Zschocke}}]{Gaia_General_2016}
{Gaia Collaboration}, {Prusti}, T., {de Bruijne}, J.~H.~J., {et~al.} 2016, \aap, 595, A1, \dodoi{10.1051/0004-6361/201629272}

\bibitem[{{Gaia Collaboration} {et~al.}(2021){Gaia Collaboration}, {Brown}, {Vallenari}, {Prusti}, {de Bruijne}, {Babusiaux}, {Biermann}, {Creevey}, {Evans}, {Eyer}, {Hutton}, {Jansen}, {Jordi}, {Klioner}, {Lammers}, {Lindegren}, {Luri}, {Mignard}, {Panem}, {Pourbaix}, {Randich}, {Sartoretti}, {Soubiran}, {Walton}, {Arenou}, {Bailer-Jones}, {Bastian}, {Cropper}, {Drimmel}, {Katz}, {Lattanzi}, {van Leeuwen}, {Bakker}, {Cacciari}, {Casta{\~n}eda}, {De Angeli}, {Ducourant}, {Fabricius}, {Fouesneau}, {Fr{\'e}mat}, {Guerra}, {Guerrier}, {Guiraud}, {Jean-Antoine Piccolo}, {Masana}, {Messineo}, {Mowlavi}, {Nicolas}, {Nienartowicz}, {Pailler}, {Panuzzo}, {Riclet}, {Roux}, {Seabroke}, {Sordo}, {Tanga}, {Th{\'e}venin}, {Gracia-Abril}, {Portell}, {Teyssier}, {Altmann}, {Andrae}, {Bellas-Velidis}, {Benson}, {Berthier}, {Blomme}, {Brugaletta}, {Burgess}, {Busso}, {Carry}, {Cellino}, {Cheek}, {Clementini}, {Damerdji}, {Davidson}, {Delchambre}, {Dell'Oro}, {Fern{\'a}ndez-Hern{\'a}ndez}, {Galluccio}, {Garc{\'\i}a-Lario},
  {Garcia-Reinaldos}, {Gonz{\'a}lez-N{\'u}{\~n}ez}, {Gosset}, {Haigron}, {Halbwachs}, {Hambly}, {Harrison}, {Hatzidimitriou}, {Heiter}, {Hern{\'a}ndez}, {Hestroffer}, {Hodgkin}, {Holl}, {Jan{\ss}en}, {Jevardat de Fombelle}, {Jordan}, {Krone-Martins}, {Lanzafame}, {L{\"o}ffler}, {Lorca}, {Manteiga}, {Marchal}, {Marrese}, {Moitinho}, {Mora}, {Muinonen}, {Osborne}, {Pancino}, {Pauwels}, {Petit}, {Recio-Blanco}, {Richards}, {Riello}, {Rimoldini}, {Robin}, {Roegiers}, {Rybizki}, {Sarro}, {Siopis}, {Smith}, {Sozzetti}, {Ulla}, {Utrilla}, {van Leeuwen}, {van Reeven}, {Abbas}, {Abreu Aramburu}, {Accart}, {Aerts}, {Aguado}, {Ajaj}, {Altavilla}, {{\'A}lvarez}, {{\'A}lvarez Cid-Fuentes}, {Alves}, {Anderson}, {Anglada Varela}, {Antoja}, {Audard}, {Baines}, {Baker}, {Balaguer-N{\'u}{\~n}ez}, {Balbinot}, {Balog}, {Barache}, {Barbato}, {Barros}, {Barstow}, {Bartolom{\'e}}, {Bassilana}, {Bauchet}, {Baudesson-Stella}, {Becciani}, {Bellazzini}, {Bernet}, {Bertone}, {Bianchi}, {Blanco-Cuaresma}, {Boch}, {Bombrun}, {Bossini},
  {Bouquillon}, {Bragaglia}, {Bramante}, {Breedt}, {Bressan}, {Brouillet}, {Bucciarelli}, {Burlacu}, {Busonero}, {Butkevich}, {Buzzi}, {Caffau}, {Cancelliere}, {C{\'a}novas}, {Cantat-Gaudin}, {Carballo}, {Carlucci}, {Carnerero}, {Carrasco}, {Casamiquela}, {Castellani}, {Castro-Ginard}, {Castro Sampol}, {Chaoul}, {Charlot}, {Chemin}, {Chiavassa}, {Cioni}, {Comoretto}, {Cooper}, {Cornez}, {Cowell}, {Crifo}, {Crosta}, {Crowley}, {Dafonte}, {Dapergolas}, {David}, {David}, {de Laverny}, {De Luise}, {De March}, {De Ridder}, {de Souza}, {de Teodoro}, {de Torres}, {del Peloso}, {del Pozo}, {Delbo}, {Delgado}, {Delgado}, {Delisle}, {Di Matteo}, {Diakite}, {Diener}, {Distefano}, {Dolding}, {Eappachen}, {Edvardsson}, {Enke}, {Esquej}, {Fabre}, {Fabrizio}, {Faigler}, {Fedorets}, {Fernique}, {Fienga}, {Figueras}, {Fouron}, {Fragkoudi}, {Fraile}, {Franke}, {Gai}, {Garabato}, {Garcia-Gutierrez}, {Garc{\'\i}a-Torres}, {Garofalo}, {Gavras}, {Gerlach}, {Geyer}, {Giacobbe}, {Gilmore}, {Girona}, {Giuffrida}, {Gomel}, {Gomez},
  {Gonzalez-Santamaria}, {Gonz{\'a}lez-Vidal}, {Granvik}, {Guti{\'e}rrez-S{\'a}nchez}, {Guy}, {Hauser}, {Haywood}, {Helmi}, {Hidalgo}, {Hilger}, {H{\l}adczuk}, {Hobbs}, {Holland}, {Huckle}, {Jasniewicz}, {Jonker}, {Juaristi Campillo}, {Julbe}, {Karbevska}, {Kervella}, {Khanna}, {Kochoska}, {Kontizas}, {Kordopatis}, {Korn}, {Kostrzewa-Rutkowska}, {Kruszy{\'n}ska}, {Lambert}, {Lanza}, {Lasne}, {Le Campion}, {Le Fustec}, {Lebreton}, {Lebzelter}, {Leccia}, {Leclerc}, {Lecoeur-Taibi}, {Liao}, {Licata}, {Lindstr{\o}m}, {Lister}, {Livanou}, {Lobel}, {Madrero Pardo}, {Managau}, {Mann}, {Marchant}, {Marconi}, {Marcos Santos}, {Marinoni}, {Marocco}, {Marshall}, {Martin Polo}, {Mart{\'\i}n-Fleitas}, {Masip}, {Massari}, {Mastrobuono-Battisti}, {Mazeh}, {McMillan}, {Messina}, {Michalik}, {Millar}, {Mints}, {Molina}, {Molinaro}, {Moln{\'a}r}, {Montegriffo}, {Mor}, {Morbidelli}, {Morel}, {Morris}, {Mulone}, {Munoz}, {Muraveva}, {Murphy}, {Musella}, {Noval}, {Ord{\'e}novic}, {Orr{\`u}}, {Osinde}, {Pagani}, {Pagano},
  {Palaversa}, {Palicio}, {Panahi}, {Pawlak}, {Pe{\~n}alosa Esteller}, {Penttil{\"a}}, {Piersimoni}, {Pineau}, {Plachy}, {Plum}, {Poggio}, {Poretti}, {Poujoulet}, {Pr{\v{s}}a}, {Pulone}, {Racero}, {Ragaini}, {Rainer}, {Raiteri}, {Rambaux}, {Ramos}, {Ramos-Lerate}, {Re Fiorentin}, {Regibo}, {Reyl{\'e}}, {Ripepi}, {Riva}, {Rixon}, {Robichon}, {Robin}, {Roelens}, {Rohrbasser}, {Romero-G{\'o}mez}, {Rowell}, {Royer}, {Rybicki}, {Sadowski}, {Sagrist{\`a} Sell{\'e}s}, {Sahlmann}, {Salgado}, {Salguero}, {Samaras}, {Sanchez Gimenez}, {Sanna}, {Santove{\~n}a}, {Sarasso}, {Schultheis}, {Sciacca}, {Segol}, {Segovia}, {S{\'e}gransan}, {Semeux}, {Shahaf}, {Siddiqui}, {Siebert}, {Siltala}, {Slezak}, {Smart}, {Solano}, {Solitro}, {Souami}, {Souchay}, {Spagna}, {Spoto}, {Steele}, {Steidelm{\"u}ller}, {Stephenson}, {S{\"u}veges}, {Szabados}, {Szegedi-Elek}, {Taris}, {Tauran}, {Taylor}, {Teixeira}, {Thuillot}, {Tonello}, {Torra}, {Torra}, {Turon}, {Unger}, {Vaillant}, {van Dillen}, {Vanel}, {Vecchiato}, {Viala}, {Vicente},
  {Voutsinas}, {Weiler}, {Wevers}, {Wyrzykowski}, {Yoldas}, {Yvard}, {Zhao}, {Zorec}, {Zucker}, {Zurbach}, \& {Zwitter}}]{2020GaiaEDR3_catalog_summary}
{Gaia Collaboration}, {Brown}, A.~G.~A., {Vallenari}, A., {et~al.} 2021, \aap, 649, A1, \dodoi{10.1051/0004-6361/202039657}

\bibitem[{Gillon {et~al.}(2017)Gillon, Triaud, Demory, Jehin, Agol, Deck, Lederer, De~Wit, Burdanov, Ingalls, {et~al.}}]{gillon2017}
Gillon, M., Triaud, A.~H., Demory, B.-O., {et~al.} 2017, Nature, 542, 456

\bibitem[{{Halbwachs} {et~al.}(2023){Halbwachs}, {Pourbaix}, {Arenou}, {Galluccio}, {Guillout}, {Bauchet}, {Marchal}, {Sadowski}, \& {Teyssier}}]{Halbwachs+Pourbaix+Arenou+etal_2023}
{Halbwachs}, J.-L., {Pourbaix}, D., {Arenou}, F., {et~al.} 2023, \aap, 674, A9, \dodoi{10.1051/0004-6361/202243969}

\bibitem[{{Hirano} {et~al.}(2011){Hirano}, {Narita}, {Shporer}, {Sato}, {Aoki}, \& {Tamura}}]{2011PASJ_hatp11b_obliquity}
{Hirano}, T., {Narita}, N., {Shporer}, A., {et~al.} 2011, \pasj, 63, 531, \dodoi{10.1093/pasj/63.sp2.S531}

\bibitem[{{Holman} \& {Murray}(2005)}]{holman2005}
{Holman}, M.~J., \& {Murray}, N.~W. 2005, Science, 307, 1288, \dodoi{10.1126/science.1107822}

\bibitem[{{Howard} {et~al.}(2010){Howard}, {Johnson}, {Marcy}, {Fischer}, {Wright}, {Bernat}, {Henry}, {Peek}, {Isaacson}, {Apps}, {Endl}, {Cochran}, {Valenti}, {Anderson}, \& {Piskunov}}]{Howard+Johnson+Marcy+etal_2010}
{Howard}, A.~W., {Johnson}, J.~A., {Marcy}, G.~W., {et~al.} 2010, \apj, 721, 1467, \dodoi{10.1088/0004-637X/721/2/1467}

\bibitem[{{Kane} \& {Torres}(2017)}]{obliquity_multiple_planets}
{Kane}, S.~R., \& {Torres}, S.~M. 2017, \aj, 154, 204, \dodoi{10.3847/1538-3881/aa8fce}

\bibitem[{{Kervella} {et~al.}(2020){Kervella}, {Arenou}, \& {Schneider}}]{Kervella+Arenou+Schneider_2020}
{Kervella}, P., {Arenou}, F., \& {Schneider}, J. 2020, \aap, 635, L14, \dodoi{10.1051/0004-6361/202037551}

\bibitem[{{Kozai}(1962)}]{kozai_1962}
{Kozai}, Y. 1962, \aj, 67, 591, \dodoi{10.1086/108790}

\bibitem[{{Li} \& {Winn}(2016)}]{Li16}
{Li}, G., \& {Winn}, J.~N. 2016, \apj, 818, 5, \dodoi{10.3847/0004-637X/818/1/5}

\bibitem[{{Li} {et~al.}(2021){Li}, {Brandt}, {Brandt}, {Dupuy}, {Michalik}, {Jensen-Clem}, {Zeng}, {Faherty}, \& {Mitra}}]{2021AJ_Li_RV_planets}
{Li}, Y., {Brandt}, T.~D., {Brandt}, G.~M., {et~al.} 2021, \aj, 162, 266, \dodoi{10.3847/1538-3881/ac27ab}

\bibitem[{{Lidov}(1962)}]{lidov_1962}
{Lidov}, M.~L. 1962, \planss, 9, 719, \dodoi{10.1016/0032-0633(62)90129-0}

\bibitem[{{Lindegren} {et~al.}(2021){Lindegren}, {Klioner}, {Hern{\'a}ndez}, {Bombrun}, {Ramos-Lerate}, {Steidelm{\"u}ller}, {Bastian}, {Biermann}, {de Torres}, {Gerlach}, {Geyer}, {Hilger}, {Hobbs}, {Lammers}, {McMillan}, {Stephenson}, {Casta{\~n}eda}, {Davidson}, {Fabricius}, {Gracia-Abril}, {Portell}, {Rowell}, {Teyssier}, {Torra}, {Bartolom{\'e}}, {Clotet}, {Garralda}, {Gonz{\'a}lez-Vidal}, {Torra}, {Abbas}, {Altmann}, {Anglada Varela}, {Balaguer-N{\'u}{\~n}ez}, {Balog}, {Barache}, {Becciani}, {Bernet}, {Bertone}, {Bianchi}, {Bouquillon}, {Brown}, {Bucciarelli}, {Busonero}, {Butkevich}, {Buzzi}, {Cancelliere}, {Carlucci}, {Charlot}, {Cioni}, {Crosta}, {Crowley}, {del Peloso}, {del Pozo}, {Drimmel}, {Esquej}, {Fienga}, {Fraile}, {Gai}, {Garcia-Reinaldos}, {Guerra}, {Hambly}, {Hauser}, {Jan{\ss}en}, {Jordan}, {Kostrzewa-Rutkowska}, {Lattanzi}, {Liao}, {Licata}, {Lister}, {L{\"o}ffler}, {Marchant}, {Masip}, {Mignard}, {Mints}, {Molina}, {Mora}, {Morbidelli}, {Murphy}, {Pagani}, {Panuzzo}, {Pe{\~n}alosa
  Esteller}, {Poggio}, {Re Fiorentin}, {Riva}, {Sagrist{\`a} Sell{\'e}s}, {Sanchez Gimenez}, {Sarasso}, {Sciacca}, {Siddiqui}, {Smart}, {Souami}, {Spagna}, {Steele}, {Taris}, {Utrilla}, {van Reeven}, \& {Vecchiato}}]{Lindegren+Klioner+Hernandez+etal_2020}
{Lindegren}, L., {Klioner}, S.~A., {Hern{\'a}ndez}, J., {et~al.} 2021, \aap, 649, A2, \dodoi{10.1051/0004-6361/202039709}

\bibitem[{{Lovis} {et~al.}(2011){Lovis}, {Dumusque}, {Santos}, {Bouchy}, {Mayor}, {Pepe}, {Queloz}, {S{\'e}gransan}, \& {Udry}}]{Lovis+Dumusque+Santos+etal_2011}
{Lovis}, C., {Dumusque}, X., {Santos}, N.~C., {et~al.} 2011, arXiv e-prints, arXiv:1107.5325, \dodoi{10.48550/arXiv.1107.5325}

\bibitem[{Lu {et~al.}(2024)Lu, An, Li, Millholland, Brandt, \& Brandt}]{lu_dynamics}
Lu, T., An, Q., Li, G., {et~al.} 2024, Planet-Planet Scattering and ZLK Migration -- The Dynamical History of HAT-P-11.
\newblock \doarXiv{2405.19511}

\bibitem[{{Mamajek} \& {Hillenbrand}(2008)}]{Mamajek+Hillenbrand_2008}
{Mamajek}, E.~E., \& {Hillenbrand}, L.~A. 2008, \apj, 687, 1264, \dodoi{10.1086/591785}

\bibitem[{{McLaughlin}(1924)}]{mcLaughlin1924some}
{McLaughlin}, D.~B. 1924, \apj, 60, 22, \dodoi{10.1086/142826}

\bibitem[{{Middelkoop}(1982)}]{Middelkoop_1982}
{Middelkoop}, F. 1982, \aap, 107, 31

\bibitem[{{Mittag} {et~al.}(2013){Mittag}, {Schmitt}, \& {Schr{\"o}der}}]{Mittag+Schmitt+Schroeder_2013}
{Mittag}, M., {Schmitt}, J.~H.~M.~M., \& {Schr{\"o}der}, K.~P. 2013, \aap, 549, A117, \dodoi{10.1051/0004-6361/201219868}

\bibitem[{{Morris} {et~al.}(2017{\natexlab{a}}){Morris}, {Hebb}, {Davenport}, {Rohn}, \& {Hawley}}]{2017-Morris-hatp11-starspots}
{Morris}, B.~M., {Hebb}, L., {Davenport}, J. R.~A., {Rohn}, G., \& {Hawley}, S.~L. 2017{\natexlab{a}}, \apj, 846, 99, \dodoi{10.3847/1538-4357/aa8555}

\bibitem[{{Morris} {et~al.}(2017{\natexlab{b}}){Morris}, {Hawley}, {Hebb}, {Sakari}, {Davenport}, {Isaacson}, {Howard}, {Montet}, \& {Agol}}]{Morris+Hawley+Hebb+etal_2017}
{Morris}, B.~M., {Hawley}, S.~L., {Hebb}, L., {et~al.} 2017{\natexlab{b}}, \apj, 848, 58, \dodoi{10.3847/1538-4357/aa8cca}

\bibitem[{{Morton} \& {Winn}(2014)}]{single_multiple_transit}
{Morton}, T.~D., \& {Winn}, J.~N. 2014, \apj, 796, 47, \dodoi{10.1088/0004-637X/796/1/47}

\bibitem[{Nagasawa \& Ida(2011)}]{nagasawa2011orbital}
Nagasawa, M., \& Ida, S. 2011, The Astrophysical Journal, 742, 72

\bibitem[{{Nesvorn{\'y}}(2018)}]{solar_disk}
{Nesvorn{\'y}}, D. 2018, \araa, 56, 137, \dodoi{10.1146/annurev-astro-081817-052028}

\bibitem[{{Noyes} {et~al.}(1984){Noyes}, {Hartmann}, {Baliunas}, {Duncan}, \& {Vaughan}}]{Noyes+Hartmann+Baliunas+etal_1984}
{Noyes}, R.~W., {Hartmann}, L.~W., {Baliunas}, S.~L., {Duncan}, D.~K., \& {Vaughan}, A.~H. 1984, \apj, 279, 763, \dodoi{10.1086/161945}

\bibitem[{Oliphant(2006)}]{numpy1}
Oliphant, T. 2006, {NumPy}: A guide to {NumPy}, USA: Trelgol Publishing.
\newblock \url{http://www.numpy.org/}

\bibitem[{{Perryman} {et~al.}(1997){Perryman}, {Lindegren}, {Kovalevsky}, {Hoeg}, {Bastian}, {Bernacca}, {Cr{\'e}z{\'e}}, {Donati}, {Grenon}, {Grewing}, {van Leeuwen}, {van der Marel}, {Mignard}, {Murray}, {Le Poole}, {Schrijver}, {Turon}, {Arenou}, {Froeschl{\'e}}, \& {Petersen}}]{HIP_TYCHO_ESA_1997}
{Perryman}, M.~A.~C., {Lindegren}, L., {Kovalevsky}, J., {et~al.} 1997, \aap, 323, L49

\bibitem[{{Petigura} {et~al.}(2017){Petigura}, {Howard}, {Marcy}, {Johnson}, {Isaacson}, {Cargile}, {Hebb}, {Fulton}, {Weiss}, {Morton}, {Winn}, {Rogers}, {Sinukoff}, {Hirsch}, \& {Crossfield}}]{2017AJ....154..107P}
{Petigura}, E.~A., {Howard}, A.~W., {Marcy}, G.~W., {et~al.} 2017, \aj, 154, 107, \dodoi{10.3847/1538-3881/aa80de}

\bibitem[{{Price-Whelan} {et~al.}(2018){Price-Whelan}, {Sip{\H{o}}cz}, {G{\"u}nther}, {Lim}, {Crawford}, {Conseil}, {Shupe}, {Craig}, {Dencheva}, {Ginsburg}, {VanderPlas}, {Bradley}, {P{\'e}rez-Su{\'a}rez}, {de Val-Borro}, {Paper Contributors}, {Aldcroft}, {Cruz}, {Robitaille}, {Tollerud}, {Coordination Committee}, {Ardelean}, {Babej}, {Bach}, {Bachetti}, {Bakanov}, {Bamford}, {Barentsen}, {Barmby}, {Baumbach}, {Berry}, {Biscani}, {Boquien}, {Bostroem}, {Bouma}, {Brammer}, {Bray}, {Breytenbach}, {Buddelmeijer}, {Burke}, {Calderone}, {Cano Rodr{\'\i}guez}, {Cara}, {Cardoso}, {Cheedella}, {Copin}, {Corrales}, {Crichton}, {D{\textquoteright}Avella}, {Deil}, {Depagne}, {Dietrich}, {Donath}, {Droettboom}, {Earl}, {Erben}, {Fabbro}, {Ferreira}, {Finethy}, {Fox}, {Garrison}, {Gibbons}, {Goldstein}, {Gommers}, {Greco}, {Greenfield}, {Groener}, {Grollier}, {Hagen}, {Hirst}, {Homeier}, {Horton}, {Hosseinzadeh}, {Hu}, {Hunkeler}, {Ivezi{\'c}}, {Jain}, {Jenness}, {Kanarek}, {Kendrew}, {Kern}, {Kerzendorf}, {Khvalko},
  {King}, {Kirkby}, {Kulkarni}, {Kumar}, {Lee}, {Lenz}, {Littlefair}, {Ma}, {Macleod}, {Mastropietro}, {McCully}, {Montagnac}, {Morris}, {Mueller}, {Mumford}, {Muna}, {Murphy}, {Nelson}, {Nguyen}, {Ninan}, {N{\"o}the}, {Ogaz}, {Oh}, {Parejko}, {Parley}, {Pascual}, {Patil}, {Patil}, {Plunkett}, {Prochaska}, {Rastogi}, {Reddy Janga}, {Sabater}, {Sakurikar}, {Seifert}, {Sherbert}, {Sherwood-Taylor}, {Shih}, {Sick}, {Silbiger}, {Singanamalla}, {Singer}, {Sladen}, {Sooley}, {Sornarajah}, {Streicher}, {Teuben}, {Thomas}, {Tremblay}, {Turner}, {Terr{\'o}n}, {van Kerkwijk}, {de la Vega}, {Watkins}, {Weaver}, {Whitmore}, {Woillez}, {Zabalza}, \& {Contributors}}]{astropy:2018}
{Price-Whelan}, A.~M., {Sip{\H{o}}cz}, B.~M., {G{\"u}nther}, H.~M., {et~al.} 2018, \aj, 156, 123, \dodoi{10.3847/1538-3881/aabc4f}

\bibitem[{{Rossiter}(1924)}]{rossiter1924detection}
{Rossiter}, R.~A. 1924, \apj, 60, 15, \dodoi{10.1086/142825}

\bibitem[{{Sanchis-Ojeda} \& {Winn}(2011)}]{2011ApJ_Sanchis-Ojeda_hatp11_star_spin}
{Sanchis-Ojeda}, R., \& {Winn}, J.~N. 2011, \apj, 743, 61, \dodoi{10.1088/0004-637X/743/1/61}

\bibitem[{Souami \& Souchay(2012)}]{souami2012solar}
Souami, D., \& Souchay, J. 2012, Astronomy \& Astrophysics, 543, A133

\bibitem[{{Stassun} {et~al.}(2017){Stassun}, {Collins}, \& {Gaudi}}]{2017AJ_hatp11_b_inclination}
{Stassun}, K.~G., {Collins}, K.~A., \& {Gaudi}, B.~S. 2017, \aj, 153, 136, \dodoi{10.3847/1538-3881/aa5df3}

\bibitem[{{Valsecchi} \& {Rasio}(2014)}]{obliquity_damp}
{Valsecchi}, F., \& {Rasio}, F.~A. 2014, \apj, 786, 102, \dodoi{10.1088/0004-637X/786/2/102}

\bibitem[{{van der Walt} {et~al.}(2011){van der Walt}, {Colbert}, \& {Varoquaux}}]{numpy2}
{van der Walt}, S., {Colbert}, S.~C., \& {Varoquaux}, G. 2011, Computing in Science and Engineering, 13, 22, \dodoi{10.1109/MCSE.2011.37}

\bibitem[{{van Leeuwen}(2007)}]{vanLeeuwen_2007}
{van Leeuwen}, F. 2007, \aap, 474, 653, \dodoi{10.1051/0004-6361:20078357}

\bibitem[{{Vaughan} {et~al.}(1978){Vaughan}, {Preston}, \& {Wilson}}]{Vaughan+Preston+Wilson_1978}
{Vaughan}, A.~H., {Preston}, G.~W., \& {Wilson}, O.~C. 1978, \pasp, 90, 267, \dodoi{10.1086/130324}

\bibitem[{{Virtanen} {et~al.}(2020){Virtanen}, {Gommers}, {Oliphant}, {Haberland}, {Reddy}, {Cournapeau}, {Burovski}, {Peterson}, {Weckesser}, {Bright}, {van der Walt}, {Brett}, {Wilson}, {Jarrod Millman}, {Mayorov}, {Nelson}, {Jones}, {Kern}, {Larson}, {Carey}, {Polat}, {Feng}, {Moore}, {Vand erPlas}, {Laxalde}, {Perktold}, {Cimrman}, {Henriksen}, {Quintero}, {Harris}, {Archibald}, {Ribeiro}, {Pedregosa}, {van Mulbregt}, \& {Contributors}}]{2020SciPy-NMeth}
{Virtanen}, P., {Gommers}, R., {Oliphant}, T.~E., {et~al.} 2020, Nature Methods, 17, 261, \dodoi{https://doi.org/10.1038/s41592-019-0686-2}

\bibitem[{{Vogt} {et~al.}(1994){Vogt}, {Allen}, {Bigelow}, {Bresee}, {Brown}, {Cantrall}, {Conrad}, {Couture}, {Delaney}, {Epps}, {Hilyard}, {Hilyard}, {Horn}, {Jern}, {Kanto}, {Keane}, {Kibrick}, {Lewis}, {Osborne}, {Pardeilhan}, {Pfister}, {Ricketts}, {Robinson}, {Stover}, {Tucker}, {Ward}, \& {Wei}}]{Vogt+Allen+Bigelow+etal_1994}
{Vogt}, S.~S., {Allen}, S.~L., {Bigelow}, B.~C., {et~al.} 1994, in \procspie, Vol. 2198, Instrumentation in Astronomy VIII, ed. D.~L. {Crawford} \& E.~R. {Craine}, 362, \dodoi{10.1117/12.176725}

\bibitem[{{von Zeipel}(1910)}]{von_zeipel_1910}
{von Zeipel}, H. 1910, Astronomische Nachrichten, 183, 345, \dodoi{10.1002/asna.19091832202}

\bibitem[{{Vousden} {et~al.}(2016){Vousden}, {Farr}, \& {Mandel}}]{Vousden+Farr+Mandel_2016}
{Vousden}, W.~D., {Farr}, W.~M., \& {Mandel}, I. 2016, \mnras, 455, 1919, \dodoi{10.1093/mnras/stv2422}

\bibitem[{{Wilson}(1963)}]{Wilson_1963}
{Wilson}, O.~C. 1963, \apj, 138, 832, \dodoi{10.1086/147689}

\bibitem[{{Wilson}(1978)}]{Wilson_1978}
---. 1978, \apj, 226, 379, \dodoi{10.1086/156618}

\bibitem[{{Winn} {et~al.}(2010){Winn}, {Johnson}, {Howard}, {Marcy}, {Isaacson}, {Shporer}, {Bakos}, {Hartman}, \& {Albrecht}}]{2010ApJ_Winn_hatp11b_obliquity}
{Winn}, J.~N., {Johnson}, J.~A., {Howard}, A.~W., {et~al.} 2010, \apjl, 723, L223, \dodoi{10.1088/2041-8205/723/2/L223}

\bibitem[{{Xuan} \& {Wyatt}(2020)}]{2020MNRAS_hatp11bc_obliquity}
{Xuan}, J.~W., \& {Wyatt}, M.~C. 2020, \mnras, 497, 2096, \dodoi{10.1093/mnras/staa2033}

\bibitem[{{Yee} {et~al.}(2018){Yee}, {Petigura}, {Fulton}, {Knutson}, {Batygin}, {Bakos}, {Hartman}, {Hirsch}, {Howard}, {Isaacson}, {Kosiarek}, {Sinukoff}, \& {Weiss}}]{2018AJ_Yee_hatp11_RVS}
{Yee}, S.~W., {Petigura}, E.~A., {Fulton}, B.~J., {et~al.} 2018, \aj, 155, 255, \dodoi{10.3847/1538-3881/aabfec}

\bibitem[{{Yee} {et~al.}(2024){Yee}, {Petigura}, {Isaacson}, {Howard}, {Blunt}, {Dalba}, {Dai}, {Fulton}, {Giacalone}, {Kane}, {Kosiarek}, {Mo{\v{c}}nik}, {Rice}, {Rubenzahl}, {Saunders}, {Tyler}, {Weiss}, \& {Zhang}}]{hatp11_RV_2024}
{Yee}, S.~W., {Petigura}, E.~A., {Isaacson}, H., {et~al.} 2024, Research Notes of the American Astronomical Society, 8, 187, \dodoi{10.3847/2515-5172/ad675e}

\bibitem[{{Zanazzi} \& {Lai}(2018)}]{zanazzi_misalignment}
{Zanazzi}, J.~J., \& {Lai}, D. 2018, \mnras, 478, 835, \dodoi{10.1093/mnras/sty1075}

\end{thebibliography}

\end{document}